\documentclass[final,3p,times,authoryear]{elsarticle}

\usepackage{amssymb}
\usepackage{amsmath}
\usepackage{multirow,booktabs,tikz,pgfplots,enumitem,bbm,amsthm}
\usetikzlibrary{arrows.meta}
\usepackage{caption,threeparttable}

\usepackage[colorlinks=true]{hyperref}

\definecolor{customblue}{RGB}{30,94,180}

\usepackage[capitalise,nameinlink]{cleveref}

\makeatletter
\newcommand{\oset}[3][0.4ex]{
	\mathrel{\mathop{#3}\limits^{
			\vbox to#1{\kern0.2\ex@
				\hbox{$\scriptstyle#2$}\vss}}}}

\newcommand{\osetd}[3][0.6ex]{
	\mathrel{\mathop{#3}\limits^{
			\vbox to#1{\kern0.2\ex@
				\hbox{$\scriptstyle#2$}\vss}}}}
\newcommand{\uset}[3][0ex]{
	\mathrel{\mathop{#3}\limits^{
			\vbox to#1{\kern5.5\ex@
				\hbox{$\scriptstyle#2$}\vss}}}}

\newcommand{\convp}{\oset{\mathrm{\scriptscriptstyle{P}\,}}{\rightarrow}}
\newcommand{\convd}{\oset{\mathrm{\scriptscriptstyle{D}\,}}{\rightarrow}}

\newcommand{\eqd}{\osetd{\mathrm{\scriptscriptstyle{D}}}{=}}

\DeclareMathOperator{\ran}{ran}

\makeatother

\theoremstyle{plain}
\newtheorem{proposition}{Proposition}[section]
\newtheorem{lemma}{Lemma}[section]

\theoremstyle{definition}
\newtheorem{assumption}{Assumption}

\crefdefaultlabelformat{#2\textup{#1}#3}

\crefname{assumption}{Assumption}{Assumptions}
\Crefname{assumption}{Assumption}{Assumptions}

\usepackage{xpatch}
\makeatletter
\xpatchcmd{\proof}
{\itshape} 
{\normalfont\bfseries}
{}{}
\xpatchcmd{\proof}
{\@addpunct{.}}
{\@addpunct{.}\enspace}
{}{}
\makeatother

\begin{document}
	
	\hypersetup{
		colorlinks=true,
		linkcolor=customblue,
		citecolor=customblue,
		urlcolor=customblue,
		filecolor=customblue,
		allcolors=customblue
	}
	
	\begin{frontmatter}
		
		\title{Modified Wilcoxon-Mann-Whitney tests of stochastic dominance}
		
		\author{Brendan K.\ Beare and Jackson D.\ Clarke}
		
		\affiliation{organization={School of Economics, University of Sydney},country={Australia}}
		
		\begin{abstract}
			Given independent samples from two univariate distributions, the one-sided Wilcoxon-Mann-Whitney statistic may be used to conduct a rank-based test of first-order stochastic dominance. We broaden the scope of applicability of such tests by showing that the bootstrap may be used to conduct valid inference in a matched pairs sampling framework permitting dependence between the two samples. Further, we show that a modified bootstrap incorporating an implicit estimate of a contact set may be used to improve power. Numerical simulations indicate that the modified bootstrap effectively controls the null rejection frequencies and delivers improved power, particularly in settings where there is strong dependence between matched pairs. We provide a brief empirical illustration involving Canadian family income data.
		\end{abstract}
		
		\begin{keyword}
			Bootstrap \sep Nonparametric approaches \sep P-P plot \sep Stochastic dominance \sep Wilcoxon-Mann-Whitney test
			
			\JEL C12 \sep C14 \sep C15
			
		\end{keyword}
		
	\end{frontmatter}
	
	\section{Introduction}
	\label{sec:intro}
	
	Tests of stochastic dominance occupy a central place in empirical economics. They are routinely employed to compare income distributions, portfolio returns, treatment effects, and many other outcomes of interest. The econometric literature on the subject begins with \citet{M89}; other foundational contributions include \citet{A96}, \citet{DD00}, \citet{BD03} and \citet{LMW05}. A major advance of central relevance to the present article was the development of a powerful bootstrap test of stochastic dominance in \citet{LSW10}. The test, called the Linton-Song-Whang test or simply the LSW test in what follows, uses a novel bootstrap procedure incorporating a preliminary estimate of a contact set. The effect of contact set estimation is to broaden the set of null configurations at which the limiting rejection rate is equal to the nominal level, thereby improving power against nearby alternatives. Subsequent literature exploring variations upon the LSW test includes \citet{DH16}, which uses a selective recentering method in place of contact set estimation; \citet{LT21}, which proposes to estimate the contact set by the method of empirical likelihood; and \citet{ZWC24}, which uses contact set estimation in conjunction with a variance-weighted Kolmogorov-Smirnov statistic.
		
	The LSW test served as a leading example in the general theory of bootstrap inference for directionally differentiable functionals developed in \citet{FS19}. The theory provides high-level conditions that guarantee the asymptotic validity of bootstrap procedures based on the estimation of a directional derivative, commonly one characterized by a contact set. The LSW test is shown to satisfy these conditions. By placing the LSW test within a more general framework of bootstrap inference based on contact set estimation, \citet{FS19} facilitated the adaptation of the central methodological insight in \citet{LSW10} to other hypothesis testing problems of interest, some only tangentially related to stochastic dominance. Examples include tests of stochastic monotonicity \citep{S18}, density ratio ordering \citep{BS19}, matrix rank \citep{CF19}, Lorenz dominance \citep{SB21}, instrument validity \citep{S23}, inverse stochastic dominance \citep{JSH24} and almost stochastic dominance \citep{SS26}.
		
	The present article contributes to this literature by developing a new rank-based test of first-order stochastic dominance. Our test statistic is the one-sided Wilcoxon-Mann-Whitney (WMW) statistic studied in \citet{ST96}. The one-sided WMW statistic can be computed from the procentile-procentile (P-P) plot for two samples drawn from two populations. The P-P plot is simply the empirical cumulative distribution function (cdf) for the first sample composed with the empirical quantile function for the second sample. When the P-P plot rises above the 45-degree line, this may be taken as evidence that the first population does not first-order stochastically dominate the second population. \cref{fig:intro} displays a P-P plot for two samples of size $10$. The one-sided WMW statistic is equal to the area shaded in red multiplied by a number depending on the two sample sizes, in this case $\sqrt{5}$. Ignoring the small white triangles above the 45-degree line---which may be understood to result from the approximation of an integral by a sum and are negligible for large sample sizes---the one-sided WMW statistic is proportional to the area below the P-P plot and above the 45-degree line, and may thus be understood to constitute an area-based measurement of the evidence against first-order stochastic dominance. It depends on the sample observations only through their pooled ranks because the P-P plot depends only on those ranks. For this reason the one-sided WMW statistic is said to be a rank-based statistic.
		
	\begin{figure}[]
		\centering	
		\begin{tikzpicture}[scale=3.5]
			\tikzstyle{vertex}=[font=\small,circle,draw,fill=yellow!20]
			\tikzstyle{edge} = [font=\scriptsize,draw,thick,-]
			\draw[black, thick] (0,0) -- (0,1);
			\draw[black, thick] (0,0) -- (1,0);
			\draw (0,0.3pt) -- (0,-1pt)
			node[anchor=north,font=\scriptsize] {$0$};
			\draw (.2,0.3pt) -- (.2,-1pt)
			node[anchor=north,font=\scriptsize] {$.2$};
			\draw (.4,0.3pt) -- (.4,-1pt)
			node[anchor=north,font=\scriptsize] {$.4$};
			\draw (.6,0.3pt) -- (.6,-1pt)
			node[anchor=north,font=\scriptsize] {$.6$};
			\draw (.8,0.3pt) -- (.8,-1pt)
			node[anchor=north,font=\scriptsize] {$.8$};
			\draw (1,0.3pt) -- (1,-1pt)
			node[anchor=north,font=\scriptsize] {$1$};
			\draw (0.3pt,0) -- (-0.8pt,0)
			node[anchor=east,font=\scriptsize] {$0$};
			\draw (0.3pt,{1/5}) -- (-0.8pt,{1/5})
			node[anchor=east,font=\scriptsize] {$.2$};
			\draw (0.3pt,{2/5}) -- (-0.8pt,{2/5})
			node[anchor=east,font=\scriptsize] {$.4$};
			\draw (0.3pt,{3/5}) -- (-0.8pt,{3/5})
			node[anchor=east,font=\scriptsize] {$.6$};
			\draw (0.3pt,{4/5}) -- (-0.8pt,{4/5})
			node[anchor=east,font=\scriptsize] {$.8$};
			\draw (0.3pt,1) -- (-0.8pt,1)
			node[anchor=east,font=\scriptsize] {$1$};
			\draw[dotted,domain=0:1]  plot(\x, \x);	
			\draw [red,-Circle](0,.1) -- (.3,.1);	
			\draw [red,dotted](.3,.1) -- (.3,.5);	
			\draw [red,-Circle](.3,.5) -- (.4,.5);	
			\draw [red,dotted](.4,.5) -- (.4,.8);	
			\draw [red,-Circle](.4,.8) -- (.7,.8);	
			\draw [red,dotted](.7,.8) -- (.7,.9);	
			\draw [red,-Circle](.7,.9) -- (.8,.9);	
			\draw [red,dotted](.8,.9) -- (.8,1);	
			\draw [red](.8,1) -- (1,1);
			\fill [red,opacity=.2] (.3,.4) rectangle (.4,.5);
			\fill [red,opacity=.2] (.4,.5) rectangle (.5,.8);
			\fill [red,opacity=.2] (.5,.6) rectangle (.6,.8);
			\fill [red,opacity=.2] (.6,.7) rectangle (.7,.8);
			\fill [red,opacity=.2] (.7,.8) rectangle (.8,.9);
			\fill [red,opacity=.2] (.8,.9) rectangle (.9,1);
		\end{tikzpicture}	
		\caption{A P-P plot for two samples of equal size $n_1=n_2=10$. The one-sided WMW statistic is equal to the red shaded area multiplied by $\sqrt{n_1n_2/(n_1+n_2)}=\sqrt{5}$.}\label{fig:intro}
		\end{figure}
		
		The asymptotic distribution of the one-sided WMW statistic when the two population distributions are equal was obtained in \citet{ST96} for cases where the two samples are drawn independently from two populations. It is the distribution of the area that lies beneath a Brownian bridge and above zero. Because this distribution is free of nuisance parameters, it is simple to implement an asymptotically valid test of first-order stochastic dominance using the one-sided WMW statistic; one need merely compare the statistic to tabulated critical values. However, in many economic applications samples are not drawn independently of one another, but rather are drawn as matched pairs from a single bivariate population distribution. In \cref{sec:WMW} we show that the asymptotic distribution of the one-sided WMW statistic under matched pairs sampling is more complicated in form than under independent sampling, and can be represented as a functional of a tied-down Brownian sheet whose covariance kernel is determined by the copula linking paired observations. The dependence of the asymptotic distribution on the unknown copula complicates inference and invalidates the use of critical values tabulated for cases where samples are independent.
		
		To overcome this obstacle we provide, in \cref{sec:bootstrap}, two bootstrap procedures for computing a critical value for the one-sided WMW statistic. The first is a standard application of the bootstrap while the second incorporates an estimate of the contact set; i.e., the set on which the population P-P curve is equal to the 45-degree line. The contact set estimator requires a user-specified tuning parameter and applies a variance-weighted exclusion rule which, in the case of matched pairs sampling, uses the empirical copula to account for dependence within pairs. \cref{prop:tauinfinity,prop:modtest} show that both bootstrap procedures result in tests whose limiting rejection rates are no greater than the nominal level at all null configurations and are equal to one at all alternative configurations. The latter result further shows that the bootstrap critical value incorporating contact set estimation delivers a limiting rejection rate equal to the nominal level everywhere on the \emph{boundary of the null}, by which we mean the set of all null configurations with positive measure contact set. We comment on the very close relationship between our procedure involving contact set estimation and the LSW test in \cref{sec:LSW}, drawing attention to some differences connected to the construction of the contact set and to the rank-based nature of the one-sided WMW statistic.
		
		The outcome of numerical simulations pertaining to the small sample performance of our bootstrap critical values, both with independent samples and with matched pairs, is reported in \cref{sec:numerical}. For both sampling frameworks the simulations show that contact set estimation can produce a large improvement in power, as is the case for the LSW test, while maintaining control of type I error for reasonable choices of the tuning parameter. The results are particularly encouraging in cases where there is strong dependence within matched pairs. \cref{sec:empirical} contains a brief empirical illustration of our procedures involving Canadian family income data. We offer some concluding thoughts in \cref{sec:final}. Mathematical proofs are provided in \cref{sec:proofs}.
		
		\section{The one-sided Wilcoxon-Mann-Whitney statistic}
		\label{sec:WMW}
		
		\subsection{Distributional assumptions and sampling frameworks}
		\label{sec:hypothesis}
		
		Let $F_1:\mathbb{R}\rightarrow[0,1]$ and $F_2:\mathbb{R}\rightarrow[0,1]$ be cdfs. Define the quantile function $Q_2:(0,1)\to\mathbb R$ by $Q_2(u)=\inf\left\{x\in\mathbb R:F_2(x)\ge u \right\}$, and define the P-P curve $R:[0,1]\to[0,1]$ by
		\begin{equation*}
			R(u)=F_1(Q_2(u))\,\,\text{for}\,\,u\in(0,1),\quad R(0)=\lim_{u\downarrow0}R(u),\quad R(1)=\lim_{u\uparrow1}R(u).
		\end{equation*}
		The P-P curve is also commonly called an ordinal dominance curve or receiver operating characteristic curve. The definition at $R(u)$ at the endpoints $u=0$ and $u=1$ guarantees that $R$ is continuous at those endpoints. Discontinuities at intermediate values of $u$ are excluded by the following assumption.
		\begin{assumption}\label{ass:distribution}
			$R$ is absolutely continuous.
		\end{assumption}
		
		Absolute continuity of $R$ is a stronger property than continuity, and implies the existence of a density $r:[0,1]\to\mathbb R$ for $R$, uniquely determined up to null sets. To be concrete we choose $r$ to be equal to the derivative of $R$ at all differentiability points of $R$, and equal to one on the zero measure set where $R$ is not differentiable. The density $r$ is nonnegative and integrable because $R$ is nondecreasing. Absolute continuity of $R$ does not require $F_1$ or $F_2$ to be continuous, but does require that any discontinuities of $F_1$ fall outside of the interior of the convex support of $F_2$.
		
		Absolute continuity of $R$ does not imply that $r$ is bounded. We wish not to exclude cases where $r$ is unbounded because they arise naturally in simple examples. For instance, if $F_1$ and $F_2$ are Gaussian cdfs with equal variances but different means $\mu_1$ and $\mu_2$ then $R$ is absolutely continuous and $r(u)$ diverges to infinity either as $u\downarrow0$ if $\mu_1<\mu_2$ or as $u\uparrow1$ if $\mu_1>\mu_2$. It has nevertheless been quite common in prior research involving P-P curves to make assumptions implying that $r$ is bounded. Examples include \citet{BM15} and \citet{BS19}, where it is assumed that $R$ is continuously differentiable.
		
		We consider two sampling frameworks: independent samples and matched pairs. Our simultaneous treatment of the two sampling frameworks follows past econometric literature including \citet{BDB14} and \citet{SB21}. In both frameworks it should be understood that $\{X_i^1\}_{i=1}^\infty$ and $\{X_i^2\}_{i=1}^\infty$ are sequences of independent and identically distributed (iid) random variables drawn from $F_1$ and $F_2$ respectively. The observed samples are $\{X_i^1\}_{i=1}^{n_1}$ and $\{X_i^2\}_{i=1}^{n_2}$. The sample sizes $n_1$ and $n_2$ should be understood to implicitly depend monotonically on an underlying index $n\in\mathbb N$, which will tend to infinity in subsequent asymptotic arguments. In both sampling frameworks we simply let $n_2=n_2(n)=n$. In the matched pairs sampling framework we also let $n_1=n_1(n)=n$ and the sample of pairs $\{(X_i^1,X_i^2)\}_{i=1}^n$ is assumed to be iid. In the independent sampling framework the iid samples $\{X_i^1\}_{i=1}^{n_1}$ and $\{X_i^2\}_{i=1}^{n_2}$ are assumed to be independent of one another, and to model the relative growth of samples sizes we assume that
		\begin{align}\label{samplesizes}
			\frac{n_1n_2}{n_1+n_2}\to\infty\quad\text{and}\quad\frac{n_2}{n_1+n_2}\to\lambda\in(0,1)\quad\text{as}\quad n\to\infty.
		\end{align}
		Note that \eqref{samplesizes} is automatically satisfied in the matched pairs sampling framework, with $\lambda=1/2$. Note also that the convergence to $\lambda\in(0,1)$ in \eqref{samplesizes} implies that $n_1n_2/(n_1+n_2)\to\infty$ if and only if $n_1\to\infty$.
		\begin{assumption}\label{ass:data} 
			Either of the following is true.
			\begin{enumerate}[label=\upshape(\roman*)]
				\item (\emph{Independent sampling.}) $\{X_i^1\}_{i=1}^\infty$ and $\{X_i^2\}_{i=1}^{\infty}$ are mutually independent sequences of iid random variables with cdfs $F_1$ and $F_2$. Furthermore, $n_1$ and $n_2$ satisfy \eqref{samplesizes}.\label{en:indep}
				\item (\emph{Matched pairs.}) $\{(X_i^1,X_i^2)\}_{i=1}^{\infty}$ is a sequence of iid pairs of random variables with marginal cdfs $F_1$ and $F_2$.\label{en:pairs}
			\end{enumerate}
		\end{assumption}
		
		The copula linking the marginal cdfs $F_1$ and $F_2$ will play an important role in the matched pairs sampling framework. In this framework we let $C:[0,1]^2\to[0,1]$ be a copula for each pair $(X_i^1,X_i^2)$. Sklar's theorem establishes the existence of $C$, and shows that $C$ is uniquely defined on the product of the closed ranges of $F_1$ and $F_2$. In the independent sampling framework we simply let $C:[0,1]^2\to[0,1]$ be the product copula $C(u,v)=uv$.
		
		\subsection{Construction of test statistic}
		\label{sec:statistic}
		
		We seek to test the null hypothesis that $F_1$ first-order stochastically dominates $F_2$ (weakly). When $R$ is continuous a simple argument shows that $F_1$ first-order stochastically dominates $F_2$ if and only if $R$ does not exceed the 45-degree line. The hypotheses we seek to discriminate between are thus
		\begin{equation*}
			\text{H}_{0}    :R(u)  \leq u  \text{ for
				all }u\in[0,1]  ;\quad
			\text{H}_{1}    :R(u)>u  \text{ for some
			}u\in[0,1]  .
		\end{equation*}
		   
		Define the empirical cdfs $\hat{F}_1:\mathbb R\to[0,1]$ and $\hat{F}_2:\mathbb R\to[0,1]$ by
		\begin{align*}
			\hat{F}_j(x)=\frac{1}{n_j}\sum_{i=1}^{n_j}
			\mathbbm{1}(X_i^j\leq x),\quad j\in\{1,2\}.
		\end{align*}
		Further define the empirical quantile function $\hat{Q}_2:(0,1)\to\mathbb R$ by $\hat{Q}_2(u)=\inf\{x\in\mathbb R:\hat{F}_2(x)\geq u\}$, and define the P-P plot $\hat{R}:[0,1]\to[0,1]$ by
		\begin{equation*}
			\hat{R}(u)=\hat{F}_1(\hat{Q}_2(u))\,\,\text{for}\,\,u\in(0,1),\quad \hat{R}(0)=\lim_{u\downarrow0}\hat{R}(u),\quad \hat{R}(1)=\lim_{u\uparrow1}\hat{R}(u).
		\end{equation*}
		Throughout this article, all notation decorated with a circumflex (or ``hat'') or with a tilde is implicitly indexed by $n$ and refers to something estimated from data.
		
		The test statistic we consider is the one-sided WMW statistic studied in \citet{ST96}. It provides an estimate of the area below the P-P curve and above the 45-degree line. Let $L^1[0,1]$ be the usual normed space of Lebesgue integrable functions from $[0,1]$ into $\mathbb R$, and define the functional $\mathcal H:L^1[0,1]\to\mathbb R$ by
		\begin{equation}\label{eq:H}
			\mathcal H(h)=\int_0^1\max\{h(u)-u,0\}\,\mathrm{d}u.
		\end{equation}
		The area below $R$ and above the 45-degree line is then $\mathcal H(R)$. We have $\mathcal H(R)=0$ under $\text{H}_0$, and $\mathcal H(R)>0$ under $\text{H}_1$. Let $X^2_{(1)},\dots,X^2_{(n_2)}$ be the sample observations $X^2_1,\dots,X^2_{n_2}$ ranked from smallest to largest. The one-sided WMW statistic $\hat{S}$ is defined by
		\begin{equation}\label{eq:WMWstatistic}
			\hat{S}=\frac{T^{1/2}_n}{n_2}\sum_{i=1}^{n_2}\max\Biggl\{\hat{F}_1\big(X_{(i)}^2\big)-\frac{i}{n_2},0\Biggr\},\quad\text{where }T_n=\frac{n_1n_2}{n_1+n_2}.
		\end{equation}
		Using the fact that $\hat{R}(u)=\hat{F}_1(X^2_{(i)})$ when $(i-1)/n_2<u\leq i/n_2$, a simple argument shows that
		\begin{equation}\label{eq:approx}
			\hat{S}\leq T^{1/2}_n\mathcal H(\hat{R})\leq\hat{S}+T_n^{1/2}/(2n_2).
		\end{equation}
		We may therefore regard $\hat{S}$ to be a convenient approximation to $\displaystyle{T_n^{1/2}\mathcal H(\hat{R})}$, with the error bound $\displaystyle{T_n^{1/2}/(2n_2)}$ vanishing asymptotically under \eqref{samplesizes}. The approximation error may be understood to correspond to the small white triangles above the 45-degree line in \cref{fig:intro}. In general there are at most $n_2$ such triangles and each has area $1/(2n_2^2)$.
		
		\subsection{Asymptotic properties}
		\label{sec:asymptotic}
		
		We will first discuss the asymptotic behavior of $\displaystyle{T_n^{1/2}(\hat{R}-R)}$, and then explain how the asymptotic behavior of $\hat{S}$ may be deduced from that of $\displaystyle{T_n^{1/2}(\hat{R}-R)}$ by applying the delta-method.
		
		\subsubsection{Convergence in distribution of the P-P process}
		\label{sec:L1odc}
		
		We call the normalized P-P plot $\displaystyle{T_n^{1/2}(\hat{R}-R)}$ the P-P process. A central ingredient to our study of the asymptotic behavior of the one-sided WMW statistic is the fact that, as $n\to\infty$, the P-P process converges in distribution in $L^1[0,1]$. Our definition of convergence in distribution is the one stated in \citet[pp.~258--259]{V98} for sequences of ``random elements'' of a metric space, in this case the separable space $L^1[0,1]$. The notation $\rightsquigarrow$ will be used to signify convergence in distribution in a metric space. If that metric space is $\mathbb R$ then we will instead write $\convd$.
		
		The limit in distribution of the P-P process may be expressed in terms of a centered Gaussian process $\mathcal B:[0,1]^2\to\mathbb R$ with covariance kernel
		\begin{align}\label{copulacov}
			\mathrm{Cov}(\mathcal B(u,v),\mathcal B(u',v'))=C(u\wedge u',v\wedge v')-C(u,v)C(u',v').
		\end{align}
		In \citet[p.~52]{GS87} the process $\mathcal B$ is referred to as a tied-down Brownian sheet with intensity measure $C$. The marginal processes $\mathcal B_1:[0,1]\to\mathbb R$ and $\mathcal B_2:[0,1]\to\mathbb R$ defined by $\mathcal B_1(u)=\mathcal B(u,1)$ and $\mathcal B_2(u)=\mathcal B(1,u)$ are Brownian bridges. The two Brownian bridges are independent if and only if $C$ is the product copula.
		
		From $\mathcal B$ we construct a centered Gaussian process $\mathcal R:[0,1]\to\mathbb R$ by setting
		\begin{equation}\label{eq:Rlim}
			\mathcal R(u)=\lambda^{1/2}\mathcal B_1(R(u))-(1-\lambda)^{1/2}r(u)\mathcal B_2(u).
		\end{equation}
		Here $\lambda$ is the limit appearing in \eqref{samplesizes}, which is equal to $1/2$ in the matched pairs sampling framework or may take any value in $(0,1)$ in the independent sampling framework. As discussed in \citet{BK26}, in the matched pairs sampling framework the distribution of $\mathcal R$ depends on $C$ only through the values taken by $C$ on the product of the closed ranges of $F_1$ and $F_2$, these values being uniquely determined by Sklar's theorem.
		
		\begin{lemma}[\citealp{BK26}]\label{prop:L1odc}
			If \cref{ass:distribution,ass:data} are satisfied then $T_n^{1/2}(\hat{R}-R)\rightsquigarrow\mathcal R$ in $L^1[0,1]$.
		\end{lemma}
		
		\cref{prop:L1odc} may be compared to, for instance, Theorem 3.1 in \citet{ACH87} for independent sampling, or Lemma 1.1 in \citet{WT21} for matched pairs sampling. Those results place stronger regularity conditions on $F_1$ and $F_2$ and establish convergence in distribution to $\mathcal R$ with respect to a uniform metric. Such convergence cannot be established under the assumptions of \cref{prop:L1odc} because these assumptions---the only relevant part being the requirement that $R$ is absolutely continuous---do not imply that $\mathcal R$ has bounded sample paths. This will not create difficulties because, as we will see, the convergence in distribution in $L^1[0,1]$ established by \cref{prop:L1odc} suffices to suitably control the asymptotic behavior of the one-sided WMW statistic.
		
		\subsubsection{Asymptotic distribution of test statistic}
		\label{sec:asymptoticWMW}
		
		In view of \eqref{eq:approx}, the difference between the test statistic $\hat{S}$ and $\displaystyle{T_n^{1/2}\mathcal H(\hat{R})}$ is asymptotically negligible. It will be more convenient for us to study the behavior of the latter quantity. The null hypothesis of first-order stochastic dominance is satisfied if and only if $\mathcal H(R)=0$. In this case we have
		\begin{align}\label{Ffdm}
			T_n^{1/2}\mathcal H(\hat R)=T_n^{1/2}(\mathcal H(\hat R)-\mathcal H(R)).
		\end{align}
		In view of the convergence in distribution of $\displaystyle{T_n^{1/2}(\hat R-R)}$ established in \cref{prop:L1odc}, we can obtain the limit distribution of $\displaystyle{T_n^{1/2}(\mathcal H(\hat R)-\mathcal H(R))}$ in \eqref{Ffdm} by applying the delta-method. Define the functional $\mathcal H'_R:L^1[0,1]\to\mathbb R$ by
		\begin{align}\label{eq:HDD}
			\mathcal H'_R(h)&=\int_{B_+}h(u)\,\mathrm{d}u+\int_{B_0}\max\{h(u),0\}\,\mathrm{d}u,
		\end{align}
		where $B_+$ and $B_0$ are the sets
		\begin{align*}
			B_+=\{u\in(0,1):R(u)>u\}\quad\text{and}\quad B_0=\{u\in(0,1):R(u)=u\}.
		\end{align*}
		Adopting the terminology introduced in \cite{LSW10}, we refer to the set $B_0$ as the contact set. It is established in \cref{lemma:HDD} that $\mathcal H'_R$ is the Hadamard directional derivative of $\mathcal H$ at $R$. See \citet{FS19} for the definition of Hadamard directional differentiability and a discussion of the delta-method oriented toward applications in econometrics. By applying the delta-method with the Hadamard directionally differentiable map $\mathcal H$ we arrive at the following consequence of \cref{prop:L1odc}.
		\begin{proposition}\label{prop:F}
			Suppose that \cref{ass:distribution,ass:data} are satisfied. Then
			\begin{equation*}
				T_n^{1/2}(\mathcal H(\hat R)-\mathcal H(R))\convd\mathcal H'_R(\mathcal R).
			\end{equation*}
			Moreover, if $\mathrm{H}_0$ is true then $\hat{S}\convd\mathcal H'_R(\mathcal R)$, whereas if $\mathrm{H}_1$ is true then $\mathrm{P}(\hat{S}>c)\to1$ for every $c\in\mathbb R$.
		\end{proposition}
		
		It is observed in \citet{ST96} that, under independent sampling and at null configurations such that $F_1=F_2$ (i.e., the least favorable case), the test statistic $\hat S$ converges in distribution to $\smallint_0^1\max\{\mathcal B(u),0\}\,\mathrm{d}u$, where $\mathcal B$ is a Brownian bridge. This follows from \cref{prop:F} by noting that, under independent sampling, if $F_1=F_2$ then $\mathcal R$ is a Brownian bridge. We may therefore construct a test of $\text{H}_0$ with limiting rejection frequency no greater than $\alpha$ at all null configurations, and equal to $\alpha$ at null configurations with $F_1=F_2$, by rejecting $\text{H}_0$ when $\hat S$ exceeds the $(1-\alpha)$-quantile of $\smallint_0^1\max\{\mathcal B(u),0\}\,\mathrm{d}u$. The $0.9$, $0.95$ and $0.99$ quantiles are reported in \citet{ST96} to be $0.39$, $0.48$ and $0.68$, respectively.
		
		In the matched pairs sampling framework $\mathcal R$ is no longer a Brownian bridge at all null configurations such that $F_1=F_2$, and depends on the unknown copula $C$. The critical values reported in \citet{ST96} therefore no longer apply. In the following section we propose a bootstrap scheme to produce critical values for $\hat S$ which apply under both independent sampling and matched pairs. Further, we show how power may be improved by incorporating an implicit estimate of the contact set into our scheme, similar to what is done in \cite{LSW10}.
		
		\section{Bootstrap procedures}
		\label{sec:bootstrap}
		
		\subsection{Construction of bootstrap critical values}
		\label{sec:bs}
		
		We consider two methods for constructing critical values. The first may be regarded as a standard implementation of the bootstrap. The second is a modification of the first based on an implicit estimate of the contact set.
		
		\subsubsection{Standard bootstrap critical values.}
		\label{sec:standardbs}
		
		 Our baseline procedure to obtain a bootstrap critical value for the test statistic $\hat S$ is as follows. We first construct bootstrap cdfs $\hat F^\ast_1:\mathbb R\to[0,1]$ and $\hat F^\ast_2:\mathbb R\to[0,1]$ by setting
		\begin{align*}
			\hat F^\ast_j(x)&=\frac{1}{n_j}\sum_{i=1}^{n_j}W^j_{i,n_j}\mathbbm{1}(X_i^j\leq x),\quad j\in\{1,2\},
		\end{align*}
		where $W^1_{n_1}=(W^1_{1,n_1},\ldots,W^1_{n_1,n_1})$ and $W^2_{n_2}=(W^2_{1,n_2},\ldots,W^2_{n_2,n_2})$ are random weights generated independently of the data. The way in which the weights are generated depends on the sampling framework. With independent samples we draw $W^1_{n_1}$ and $W^2_{n_2}$ independently of one another from the multinomial distribution with equal probabilities over the categories $1,\ldots,n_1$ and $1,\ldots,n_2$ respectively. With matched pairs we draw $W^1_n$ from the multinomial distribution with equal probabilities over the categories $1,\ldots,n$, and then set $W^2_n=W^1_n$. In either sampling framework, we then construct the bootstrap quantile function $\hat{Q}^\ast_2:(0,1)\to\mathbb R$ by setting $\hat Q^\ast_2(u)=\inf\{x\in\mathbb R:\hat F^\ast_2(x)\geq u\}$, and the bootstrap P-P plot $\hat R^\ast:[0,1]\to[0,1]$ by setting
		\begin{equation*}
			\hat{R}^\ast(u)=\hat{F}_1^\ast(\hat{Q}_2^\ast(u))\,\,\text{for}\,\,u\in(0,1),\quad \hat{R}^\ast(0)=\lim_{u\downarrow0}\hat{R}^\ast(u),\quad \hat{R}^\ast(1)=\lim_{u\uparrow1}\hat{R}^\ast(u).
		\end{equation*}
		We then compute the bootstrap test statistic
		\begin{align}\label{eq:bsstat}
			\hat S^\ast&=\frac{T_n^{1/2}}{n_2}\sum_{i=1}^{n_2}\max\Biggl\{\hat{R}^\ast\biggl(\frac{i}{n_2}\biggr)-\hat R\biggl(\frac{i}{n_2}\biggr),0\Biggr\}.
		\end{align}
		To obtain a test with nominal level $\alpha$ we independently generate a large number $N$ of bootstrap test statistics and choose as our critical value the $\lceil N(1-\alpha)\rceil$-th smallest of these, where $\lceil\cdot\rceil$ rounds up to the nearest integer.
		
		\subsubsection{Modified bootstrap critical values.}
		\label{sec:modbs}
		
		Our second bootstrap procedure involves modifying the bootstrap statistic defined in \eqref{eq:bsstat} so as to incorporate an implicit estimate of $B_0$, the contact set. The first step in this procedure is to compute, for $i\in\{1,\dots,n_2\}$, an estimate $\hat{V}_i$ of the variance of $\displaystyle{T_n^{1/2}\hat R(i/n_2)}$. This is discussed in more detail below. Next, to generate a single bootstrap test statistic, we generate $\hat{R}^\ast$ in the same way as in the standard bootstrap procedure described in \cref{sec:standardbs}, and then compute the modified bootstrap test statistic
		\begin{equation}\label{eq:bsstatmod}
			\tilde S^\ast=\frac{T_n^{1/2}}{n_2}\sum_{i=1}^{n_2}\max\Biggl\{\hat{R}^\ast\biggl(\frac{i}{n_2}\biggr)-\hat R\biggl(\frac{i}{n_2}\biggr),0\Biggr\}\,\mathbbm{1}\Biggl(T_n^{1/2}\biggl(\hat R\biggl(\frac{i}{n_2}\biggr)-\frac{i}{n_2}\biggr)> -\tau_n\hat{V}^{1/2}_i\Biggr)
		\end{equation}
		where $\tau_n\in(0,\infty)$ is a tuning parameter. Note that if we set $\tau_n=\infty$ in \eqref{eq:bsstatmod} then, ignoring the possibility that $\hat{V}_i=0$ for some $i$, we recover the standard bootstrap test statistic defined in \eqref{eq:bsstat}. To obtain a test with nominal level $\alpha$ we independently generate a large number $N$ of modified bootstrap test statistics and choose as our critical value the $\lceil N(1-\alpha)\rceil$-th smallest of these. We reject $\text{H}_0$ when $\hat S$ exceeds this critical value.
		
		The role of the indicator function in \eqref{eq:bsstatmod} is to exclude summands for which $\hat R(i/n_2)$ falls below $i/n_2$ by at least $\tau_n$ estimated standard deviations. It provides an implicit estimate of the contact set. This will be made more clear in \cref{sec:modbsasy}. In the development of asymptotics to follow we will assume that $\tau_n$ diverges to infinity at a controlled rate as $n\to\infty$. See \cref{ass:tau} below. The numerical simulations reported in \cref{sec:numerical} may be used to guide the choice of $\tau_n$ in practice.
		
		The estimators $\hat{V}_i$ may be chosen to approximate pointwise variances of $\mathcal R$. It suffices for our purposes to focus on estimators that work well on the contact set $B_0$. If we set $R(u)=u$ and $r(u)=1$ in \eqref{eq:Rlim} then, by working with the covariance kernel in \eqref{copulacov}, we find that $\mathrm{Var}(\mathcal R(u))=u-C(u,u)$. We therefore propose setting
		\begin{equation*}
			\hat{V}_i=\begin{cases}\frac{i}{n_2}-\frac{i^2}{n_2^2}&\text{with independent samples}\\
				\frac{i}{n}-\hat{C}\bigl(\frac{i}{n},\frac{i}{n}\bigr)&\text{with matched pairs,}\end{cases}
		\end{equation*}
		where with matched pairs we define $\hat{C}:[0,1]^2\to[0,1]$ to be the empirical copula
		\begin{align*}
			\hat{C}(u,v)&=\frac{1}{n}\sum_{i=1}^n\mathbbm{1}\Bigl(\hat{F}_1(X_i^1)\leq u,\hat{F}_2(X_i^2)\leq v\Bigr).
		\end{align*}
		
		\subsection{Asymptotic properties}
		\label{sec:bsasym}
		
		\subsubsection{Bootstrap approximation of the P-P process.}
		\label{sec:PPbsasy}
		
		To study the asymptotic behavior of our bootstrap procedures we require a bootstrap analogue to the convergence in distribution of the P-P process established in \cref{prop:L1odc}. See \citet[p.~333]{V98} for the definition of convergence in distribution conditional on the data in probability.
		\begin{lemma}[\citealp{BK26}]\label{prop:bswc}
			If \cref{ass:distribution,ass:data} are satisfied then $T_n^{1/2}(\hat{R}^\ast-\hat{R})\rightsquigarrow\mathcal R$ in $L^1[0,1]$ conditional on the data in probability.
		\end{lemma}
		
		\cref{prop:bswc} may be roughly understood to mean that, for large $n$, the distribution of $\displaystyle{T_{n}^{1/2}}(\hat{R}^\ast-\hat{R})$ conditional on the data is, with high probability, close to the distribution of $\mathcal R$. This is useful to know because the distribution of $\displaystyle{T_{n}^{1/2}(\hat{R}^\ast-\hat{R})}$ conditional on the data is precisely what we simulate by bootstrapping.
		
		\subsubsection{Standard bootstrap critical values.}
		\label{sec:stdbsasy}
		
		Let $I:[0,1]\to[0,1]$ be the identity map, and recall the definition of $\mathcal H'_R$ given in \eqref{eq:HDD}. The standard bootstrap test statistic $\hat{S}^\ast$ defined in \eqref{eq:bsstat} satisfies
		\begin{equation}\label{eq:bSint2}
			\hat S^\ast=\int_0^1\max\big\{T_n^{1/2}(\hat R^\ast(u)-\hat R(u)),0\big\}\,\mathrm{d}u=\mathcal H'_{I}(\displaystyle{T_n^{1/2}(\hat{R}^\ast-\hat{R})}).
		\end{equation}
		If $\mathrm{H}_0$ is true then $\mathcal H'_R\leq\mathcal H'_{I}$. Thus
		\begin{equation*}
			\hat S^\ast\geq\mathcal H'_{R}(\displaystyle{T_n^{1/2}(\hat{R}^\ast-\hat{R})})\quad\text{if }\mathrm{H}_0\text{ is true, with equality if }F_1=F_2.
		\end{equation*}
		
		In view of \cref{prop:bswc} we may expect that the distribution of the quantity on the right-hand side of the last inequality conditional on the data is close to the distribution of $\mathcal H'_R(\mathcal R)$ for large $n$. We know from \cref{prop:L1odc} that $\hat{S}$ converges in distribution to $\mathcal H'_R(\mathcal R)$ under $\mathrm{H}_0$. This suggests that if $\mathrm{H}_0$ is true then a reasonable upper bound on the quantiles of $\hat{S}$ may be provided by the corresponding quantiles of $\hat{S}^\ast$ conditional on the data. On the other hand, if $\mathrm{H}_1$ is true then we know from \cref{prop:F} that $\hat{S}$ diverges in probability to infinity as $n$ grows, whereas \cref{prop:bswc} and \eqref{eq:bSint2} together suggest that the quantiles of $\hat{S}^\ast$ conditional on the data ought to become close to the corresponding quantiles of $\mathcal H'_{I}(\mathcal R)$.
		
		The loose reasoning provided in the previous paragraph provides a heuristic justification for rejecting $\mathrm{H}_0$ when $\hat{S}$ exceeds the standard bootstrap critical value computed as described in \cref{sec:standardbs}. To provide a rigorous justification we will need to exclude certain degenerate cases. The next assumption serves this purpose.
		
		\begin{assumption}\label{ass:nondegenerate}
			$R$ is not identically equal to zero or one, and $C(u,u)<u$ for a.e.\ $u\in\ran(F_1)\cap\ran(F_2)$.
		\end{assumption}
		
		The assumption on $R$ excludes cases where $F_1$ assigns all mass either entirely to the right or entirely to the left of the convex support of $F_2$. The assumption on $C$ excludes cases where there is extreme positive dependence within matched pairs. Note that the Fr\'{e}chet-Hoeffding upper bound for $C(u,u)$ is $u$.
		
		\begin{proposition}\label{prop:tauinfinity}
			Suppose that \cref{ass:distribution,ass:data,ass:nondegenerate} are satisfied. Let $\alpha\in(0,1/2)$ and let $\hat{c}_{1-\alpha}$ be the $(1-\alpha)$-quantile of $\hat{S}^\ast$ conditional on the data; i.e.,
			\begin{equation*}
				\hat{c}_{1-\alpha}=\inf\bigg\{c\in\mathbb R: \mathrm{P}\,\Big(   \hat S^\ast\le c\bigm|   \{  X_{i}^{1}\}  _{i=1}^{n_{1}}  ,\{  X_{i}^{2}\}  _{i=1}^{n_{2}}      \Big) \ge 1-\alpha   \bigg\}.
			\end{equation*}
			\begin{enumerate}[label=\upshape(\roman*)]
				\item If $F_1=F_2$ then $\mathrm P(\hat S>\hat{c}_{1-\alpha})\to\alpha$.
				\item If \(\mathrm H_{0}\) is true then $\lim\sup\mathrm P(\hat S>\hat{c}_{1-\alpha})\leq\alpha$.
				\item If \(\mathrm H_{1}\) is true then $\mathrm P(\hat S>\hat{c}_{1-\alpha})\to1$.
			\end{enumerate}
		\end{proposition}
		
		\subsubsection{Modified bootstrap critical values.}
		\label{sec:modbsasy}
		
		To study our modified bootstrap procedure we adopt an asymptotic framework in which the tuning parameter $\tau_n$ is assumed to diverge to infinity at a controlled rate as $n\to\infty$.
		\begin{assumption}\label{ass:tau}
			$\tau_n\to\infty$ and $\displaystyle{T_n^{-1/2}\tau_n\to0}$ as $n\to\infty$.
		\end{assumption}
		We claimed in \cref{sec:modbs} that the indicator function in \eqref{eq:bsstatmod} has the effect of providing an implicit estimate of the contact set $B_0$. The implicit estimate we were referring to is
		\begin{equation}\label{eq:WMWcs}
			\hat{B}_0=\Big\{  u\in(0,1)  :T_n^{1/2}(\hat{R}(u)-u)>-\tau_{n}\hat{V}_{\lceil n_2u\rceil}^{1/2}\Big\},
		\end{equation}
		though we note that under $\mathrm{H}_1$ it is more natural to regard $\hat{B}_0$ as an estimate of $B_0\cup B_+$. Of course $B_+$ is empty under $\mathrm{H}_0$. Using $\hat{B}_0$ we define the data-dependent functional $\hat{\mathcal H}':L^1[0,1]\to\mathbb R$ by
		\begin{align*}
			\hat{\mathcal H}'(h)&=\int_{\hat{B}_0}\max\{h(u),0\}\,\mathrm{d}u.
		\end{align*}
		The functional $\hat{\mathcal H}'$ may be viewed as an implicit estimate of the directional derivative $\mathcal H'_R$ defined in \eqref{eq:HDD}, again noting that $B_+$ is empty under $\mathrm{H}_0$. The modified bootstrap test statistic $\tilde S^\ast$ defined in \eqref{eq:bsstatmod} may be rewritten as
		\begin{align}\label{eq:implicit}
			\tilde S^\ast&=\hat{\mathcal H}'(T_n^{1/2}(\hat R^\ast-\hat R)),
		\end{align}
		revealing the connection between $\tilde S^\ast$ and the implicitly estimated contact set.
		
		The representation of $\tilde{S}^\ast$ given in \eqref{eq:implicit} is useful because it facilitates the application of results in \citet{FS19} providing conditions sufficient for the validity of modified bootstrap procedures. By showing that $\hat{\mathcal H}'$ suitably approximates $\mathcal H'_R$ under $\mathrm{H}_0$, we are able to use \cref{prop:bswc} above and Theorem 3.2 in \citet{FS19} to show that if $\mathrm{H}_0$ is true then, for large $n$, the distribution of $\displaystyle{\hat{\mathcal H}'(T_n^{1/2}(\hat R^\ast-\hat R))}$ conditional on the data is, with high probability, close to the distribution of $\mathcal H'_R(\mathcal R)$. See \cref{lem:FS} for a precise statement. We use \cref{lem:FS} to establish the following result.
		\begin{proposition}\label{prop:modtest}
			Suppose that \cref{ass:distribution,ass:data,ass:nondegenerate,ass:tau} are satisfied. Let $\alpha\in(0,1/2)$ and let $\tilde{c}_{1-\alpha}$ be the $(1-\alpha)$-quantile of $\tilde{S}^\ast$ conditional on the data; i.e.,
			\begin{equation*}
				\tilde{c}_{1-\alpha}=\inf\bigg\{c\in\mathbb R: \mathrm{P}\,\Big(   \tilde S^\ast\le c\bigm|   \{  X_{i}^{1}\}  _{i=1}^{n_{1}}  ,\{  X_{i}^{2}\}  _{i=1}^{n_{2}}      \Big) \ge 1-\alpha   \bigg\}.
			\end{equation*}
			\begin{enumerate}[label=\upshape(\roman*)]
				\item If \(\mathrm H_{0}\) is true and $R(u)=u$ on a set of positive measure then
				\begin{align*}
					\lim_{\eta\downarrow0}\lim_{n\to\infty}\mathrm P(\hat{S}>\max\{\tilde{c}_{1-\alpha},\eta\})=\lim_{n\to\infty}\mathrm P(\hat{S}>\tilde{c}_{1-\alpha})=\alpha.
				\end{align*}
				\item If \(\mathrm H_{0}\) is true and $R(u)<u$ a.e.\ then $\mathrm P(\hat{S}>\max\{\tilde{c}_{1-\alpha},\eta\})\to0$ for each $\eta>0$.
				\item If \(\mathrm H_{1}\) is true then $\mathrm P(\hat S>\max\{\tilde{c}_{1-\alpha},\eta\})\to1$ for each $\eta\geq0$.
			\end{enumerate}
		\end{proposition}
		
		The role of the constant $\eta$ appearing in the statement of \cref{prop:modtest} is to control the limiting rejection frequency at null configurations with zero measure contact set, i.e.\ case (ii). At such configurations both the test statistic $\hat S$ and modified bootstrap critical value $\tilde c_{1-\alpha}$ converge in probability to zero, making it difficult to characterize the rejection frequency within our first-order asymptotic framework. See \citet[pp.\ 559--560]{DH16}, \citet[pp.\ 15--16]{BS19} and \citet[p.\ 195]{SB21} for further discussion of this issue. Also see \citet{LSW10}, where the regularity condition introduced in Definition 3 plays a similar role to $\eta$ by excluding null configurations at which the critical value converges in probability to zero. In \citet{DH16} it is recommended to set $\eta$ equal to a very small value such as $10^{-6}$ in practice, and to use $\max\{\tilde c_{1-\alpha},\eta\}$ as a critical value rather than $\tilde c_{1-\alpha}$; however it is also reported that in numerical simulations there is no difference between setting $\eta=10^{-6}$ and $\eta=0$. We have found the same in numerical simulations and on this basis recommend setting $\eta=0$.
		
		\section{Relation to the Linton-Song-Whang test}
		\label{sec:LSW}
		
		Our procedure for modifying bootstrap critical values closely resembles the implementation of the bootstrap in the Linton-Song-Whang (LSW) test of stochastic dominance, proposed in \citet{LSW10}. The discussion therein focuses on the matched pairs sampling framework, and our discussion in this section will do the same, though we note that the LSW test may also be applied, with obvious modifications, in a setting with independent differently-sized samples. The LSW test statistic for the null hypothesis of first-order stochastic dominance is
		\begin{equation}\label{eq:LSWstatistic}
			\hat{L}=\int_{-\infty}^\infty\max\big\{\sqrt{n}(\hat{F}_1(x)-\hat{F}_2(x)),0\big\}^pw(x)\,\mathrm{d}x,
		\end{equation}
		where $p=2$ and where $w$ is a user-specified weight function. To test higher-order stochastic dominance the empirical cdfs $\hat{F}_1$ and $\hat{F}_2$ are replaced with cumulative integrals thereof, and $w$ should be chosen to ensure that $\hat{L}$ is finite. To test first-order stochastic dominance one may simply set $w\equiv 1$; see \citet[p.~74]{W19}. However if one were to replace $w(x)\,\mathrm{d}x$ with $\mathrm{d}\hat{F}_2(x)$ in \eqref{eq:LSWstatistic} then the resulting statistic with $p=2$ is a one-sided Cram\'{e}r-von Mises statistic. Moreover, if we instead set $p=1$ and if there are no ties in the observations $X_1^2,\dots,X_n^2$ then
		\begin{equation*}
			\int_{-\infty}^\infty\max\big\{\sqrt{n}(\hat{F}_1(x)-\hat{F}_2(x)),0\big\}\,\mathrm{d}\hat{F}_2(x)=\frac{1}{\sqrt{n}}\sum_{i=1}^n\max\Biggl\{\hat{R}\,\bigg(\frac{i}{n}\bigg)-\frac{i}{n},0\Biggr\}=\sqrt{2}\,\hat{S};
		\end{equation*}
		see \eqref{eq:WMWstatistic}. Thus, in the absence of tied observations, the one-sided WMW statistic (scaled by $\sqrt{2}$) is obtained by modifying the definition of the LSW test statistic in \eqref{eq:LSWstatistic} so that $p=1$ and so that $w(x)\,\mathrm{d}x$ is replaced by $\mathrm{d}\hat{F}_2(x)$.
		
		The contact set relevant for the LSW test (of first-order stochastic dominance) is the set $D_0\subseteq\mathbb R$ comprised of all $x$ such that $F_1(x)=F_2(x)$. The LSW test relies on an estimate of $D_0$ given by
		\begin{equation}\label{eq:LSWcs}
			\hat{D}_0=\big\{x\in\mathbb R:|\hat{F}_1(x)-\hat{F}_2(x)|<\kappa_n\big\},
		\end{equation}
		where $\kappa_n$ is a tuning parameter chosen such that $\kappa_n\to0$ and $\sqrt{n}\,\kappa_n\to\infty$ as $n\to\infty$. The critical value for the LSW test is computed from the simulated distribution, conditional on the data, of the bootstrap statistic $\tilde{L}^\ast$ defined by
		\begin{equation}\label{eq:LSWbs}
			\tilde{L}^\ast=\int_{\hat{D}_0}\max\big\{\sqrt{n}(\hat{F}^\ast_1(x)-\hat{F}^\ast_2(x))-\sqrt{n}(\hat{F}_1(x)-\hat{F}_2(x)),0\big\}^2w(x)\,\mathrm{d}x.
		\end{equation}
		A comparable expression for the modified bootstrap statistic $\tilde{S}^\ast$ (again scaled by $\sqrt{2}$) is
		\begin{equation*}
			\sqrt{2}\,\tilde{S}^\ast=\int_{\hat{B}_0}\max\big\{\sqrt{n}(\hat{R}^\ast(u)-\hat{R}(u)),0\big\}\,\mathrm{d}u;
		\end{equation*}
		see \eqref{eq:implicit}. There is thus a close connection between the modified WMW test and the LSW test.
		
		Let $G$ be the set of all monotone bijections $g:\mathbb R\to\mathbb R$. The property of first-order stochastic dominance is invariant under $G$ in the following sense: if $X^1$ and $X^2$ are random variables such that $X^1$ first-order stochastically dominates $X^2$, then $g(X^1)$ first-order stochastically dominates $g(X^2)$ for every $g\in G$. Consider the effect on the LSW test and on the modified WMW test of applying some $g\in G$ to all of the sample observations $X_i^1$ and $X_i^2$. The outcome of the modified WMW test is unaffected by the application of $g$ because $\hat{R}$, $\hat{R}^\ast$ and $\hat{C}$ (the last of which is used to construct the variance estimates $\hat{V}_i$) all depend on the sample observations only through their pooled ranks. This property is not shared by the LSW test due to the use of the fixed measure $w(x)\,\mathrm{d}x$ rather than the empirical measure $\mathrm{d}\hat{F}_2(x)$ in \eqref{eq:LSWstatistic} and \eqref{eq:LSWbs}. Applying a transformation $g\in G$ to all observations can change the outcome of the LSW test from rejection to non-rejection, or vice-versa. We illustrate this phenomenon in the empirical application reported in \cref{sec:empirical}. The non-invariance of the LSW test of first-order stochastic dominance to transformations of the data in $G$ means that the test does not satisfy Lehmann's \emph{principle of invariance}---see \citet[pp.~11--13, 241--243]{LR22}---and so may be susceptible to manipulation by unscrupulous practitioners. This concern is relevant only when testing first-order stochastic dominance, as higher orders of stochastic dominance are not invariant under $G$.
		
		The estimated contact set $\hat{D}_0$ used to implement the LSW test is comprised of those points $x\in\mathbb R$ such that $\hat{F}_1(x)-\hat{F}_2(x)$ is \emph{between} the \emph{uniform} thresholds $\pm\,\kappa_n$; see \eqref{eq:LSWcs}. The corresponding set $\hat{B}_0$ defined in \eqref{eq:WMWcs} is comprised of those points $u\in(0,1)$ such that $T_n^{1/2}(\hat{R}(u)-u)$ is \emph{above} the \emph{variance-weighted} threshold $-\tau_n\hat{V}^{1/2}_{\lceil n_2 u\rceil}$. An alternative definition of the modified bootstrap statistic $\tilde{S}^\ast$ that uses two-sided uniform thresholding as in the LSW test is
		\begin{equation}\label{eq:bsstatmod2}
			\tilde S^\ast=\frac{T_n^{1/2}}{n_2}\sum_{i=1}^{n_2}\max\Biggl\{\hat{R}^\ast\biggl(\frac{i}{n_2}\biggr)-\hat R\biggl(\frac{i}{n_2}\biggr),0\Biggr\}\,\mathbbm{1}\Biggl(T_n^{1/2}\bigg|\hat R\biggl(\frac{i}{n_2}\biggr)-\frac{i}{n_2}\bigg|< \tau_n\Biggr),
		\end{equation}
		which should be compared to our preferred definition of $\tilde{S}^\ast$ in \eqref{eq:bsstatmod}. Close inspection of the proof of \cref{prop:modtest} in \cref{sec:proofs} shows that this result remains valid under the alternative definition of $\tilde{S}^\ast$, with only minor adjustments needed in the proof of \cref{lem:FS}. We do not have a good theoretical justification for preferring \eqref{eq:bsstatmod} to \eqref{eq:bsstatmod2}, but recommend using \eqref{eq:bsstatmod} on the basis of unreported numerical simulations. It may be surprising that the indicator function in \eqref{eq:bsstatmod} is not chosen to exclude summands for which $T_n^{1/2}(\hat{R}(i/n_2)-i/n_2)\geq\tau_n \hat{V}_{i}^{1/2}$, as doing so would seem to mechanically improve power. However, if such summands are excluded then for each fixed $n$ a larger value of $\tau_n$ is needed to suitably control the rate of type I error, leaving the overall effect on power ambiguous.

		\section{Numerical simulations}
		\label{sec:numerical}
		
		To investigate the small sample properties of the modified WMW test we ran a number of Monte Carlo simulations. In each simulation we used $10^5$ Monte Carlo repetitions to compute rejection frequencies. In each of these repetitions we randomly generated an iid sample of pairs $\{(X_i^1,X_i^2)\}_{i=1}^{n}$, with $n$ varying from $25$ to $1000$ as described below. The samples were generated with $F_1$ normalized to be the uniform distribution on $[0,1]$, with $F_2$ chosen to obtain a desired P-P curve $R$ as described below, and with $C$ chosen to be the Gaussian copula with correlation parameter $\rho$ equal to $0$, $.25$, $.5$ or $.75$. Setting $\rho=0$ places us in the independent sampling framework with equally-sized samples, while setting $\rho>0$ places us in the matched pairs sampling framework with positive dependence within pairs. Bootstrap critical values were computed using $10^3$ bootstrap samples.
		
		To provide a point of comparison we also report rejection frequencies obtained using the test of first-order stochastic dominance proposed in \citet{DH16}. See \citet[pp.~80--85]{W19} for a succinct treatment. The Donald-Hsu (DH) test has asymptotic properties comparable to those established for the modified WMW test in \cref{prop:modtest}. However the DH test is based on the one-sided Kolmogorov-Smirnov statistic rather than the one-sided WMW statistic, and uses a selective recentering method to modify bootstrap critical values rather than a contact set estimator. We use the DH test in our simulations rather than the LSW test because the non-invariance of the latter test to strictly increasing transformations of the data makes it easy to manipulate the design of simulations so that the test appears more or less powerful. Like the modified WMW test, the DH test of first-order stochastic dominance depends on the sample observations only through their pooled ranks.
		
		\subsection{Null rejection frequencies}
		\label{sec:numericalnull}
		
		\subsubsection{Independent sampling framework}
		\label{sec:numericalnullindep}
		
		In \cref{lfctab} we report rejection frequencies at the least favorable case $R(u)=u$, i.e.\ $F_1=F_2$, for the modified WMW test and for the DH test with independent equally-sized samples. Rejection frequencies are reported at the nominal levels $\alpha=.05,.01$ and for sample sizes $n=25,50,100,200,500,1000$. The tuning parameter $\tau_n$ for the modified WMW test was set equal to the values $.5,.75,1,1.25,1.5,\infty$, with the value $\infty$ corresponding to the standard bootstrap critical value described in \cref{sec:standardbs}. The tuning parameter for the DH test was set equal to the values $-.025,-.05,-.1,-.15,-.2,-\infty$, with the value $-\infty$ corresponding to one of the standard bootstrap procedures described in \citet{BD03}. Note that the simulations reported in \citet{DH16} use a tuning parameter value of $-.1\sqrt{\log\log(n_1+n_2)}$, which decreases from $-.117$ to $-.142$ as the two equal sample sizes increase from $25$ to $1000$, thus falling within the range of tuning parameter values considered here.
		
		\cref{lfctab} shows that the modified WMW test delivers a rejection frequency that is generally close to, but slightly less than, the nominal level. Some over-rejection is observed with the smallest sample sizes and tuning parameters. On the other hand, the DH test delivers a rejection frequency that is generally close to, but slightly greater than, the nominal level. As expected, the rejection frequencies rise as the tuning parameters decrease in magnitude.
		
		\begin{table}[]
			\centering
			\begin{threeparttable}
				\caption{Rejection frequencies at the least favorable case with independent equally-sized samples.}\label{lfctab}
				{\small
				\begin{tabular}{rrcccccccccccc}
					\toprule
					\multicolumn{2}{c}{}&\multicolumn{6}{c}{Mod.\ Wilcoxon-Mann-Whitney}&\multicolumn{6}{c}{Donald-Hsu}\\
					\cmidrule(lr){3-8}\cmidrule(lr){9-14}
					\multicolumn{2}{r}{Tun.\ par.} & .5 & .75 & 1 & 1.25 & 1.5 & $\infty$ & -.025 & -.05 & -.1 & -.15 & -.2 & $-\infty$\\
					\midrule
					$\alpha$&$n$&&&&&&&&&&&\\
					\midrule
					\multirow{6}{*}{.05}& 25 &5.6&4.6&4.2&4.0&3.9&3.8&7.2&7.2&7.2&7.2&7.2&6.9\\
					& 50 &4.9&4.4&4.2&4.1&4.0&4.0&6.1&6.1&6.1&6.0&6.0&5.9\\
					& 100 &4.8&4.5&4.4&4.4&4.3&4.3&6.3&6.3&6.3&6.2&6.2&6.1\\
					& 200 &4.7&4.4&4.3&4.3&4.2&4.2&5.5&5.5&5.5&5.5&5.5&5.3\\
					& 500 &4.7&4.4&4.3&4.2&4.2&4.2&5.2&5.2&5.2&5.2&5.2&5.1\\
					& 1000 &4.7&4.4&4.3&4.2&4.2&4.2&5.4&5.4&5.4&5.4&5.3&5.2\\
					
					\midrule
					\multirow{6}{*}{.01}& 25 &2.1&1.6&1.4&1.3&1.3&1.2&1.9&1.9&1.9&1.9&1.9&1.9\\
					& 50 &1.3&1.1&1.1&1.1&1.1&1.0&1.4&1.4&1.4&1.4&1.4&1.4\\
					& 100 &1.1&1.0&1.0&1.0&1.0&1.0&1.2&1.2&1.2&1.2&1.2&1.2\\
					& 200 &0.9&0.9&0.9&0.9&0.9&0.9&1.2&1.2&1.2&1.2&1.2&1.2\\
					& 500 &1.0&0.9&0.9&0.9&0.9&0.9&1.0&1.0&1.0&1.0&1.0&1.0\\
					& 1000 &0.9&0.9&0.9&0.9&0.9&0.8&1.0&1.0&1.0&1.0&1.0&1.0\\
					\bottomrule
				\end{tabular}
				}
			\end{threeparttable}
		\end{table}
		
		In \cref{fig:sizeinterior} we report rejection frequencies obtained with independent samples of size $n=500$ and nominal level $\alpha=.05$ for two parametric families of P-P curves satisfying the null hypothesis of first-order stochastic dominance. The top-left and bottom-left panels of \cref{fig:sizeinterior} display the two families of P-P curves. The family in the top-left has the parametrization $R_\gamma(u)=u^{1+\gamma}$ with $\gamma\geq0$, and the family in the bottom-left has the parametrization
		\begin{align*}
			R_\gamma(u)&=\begin{cases}\Phi\bigl(\mathrm{e}^\gamma\Phi^{-1}(u)\bigr)&\text{for }u\in(0,.5)\\u&\text{for }u\in[.5,1)\end{cases}
		\end{align*}
		with $\gamma\geq0$, where $\Phi$ is the standard normal cdf and $\Phi^{-1}$ the corresponding quantile function. Note that for the bottom-left family the contact set is always $[.5,1)$ when $\gamma>0$, so that we are on the boundary of the null, whereas for the top-left family the contact set is empty when $\gamma>0$, so that we are in the interior of the null. In both families we obtain the least favorable case when $\gamma=0$.
		
		\begin{figure}
			\centering
			\begin{tikzpicture}[scale=3.5]
				\tikzstyle{vertex}=[font=\small,circle,draw,fill=yellow!20]
				\tikzstyle{edge} = [font=\scriptsize,draw,thick,-]
				\draw[black, thick] (0,0) -- (0,1);
				\draw[black, thick] (0,0) -- (1,0);
				\draw (0,0.3pt) -- (0,-1pt)
				node[anchor=north,font=\scriptsize] {$0$};
				\draw (.2,0.3pt) -- (.2,-1pt)
				node[anchor=north,font=\scriptsize] {$.2$};
				\draw (.4,0.3pt) -- (.4,-1pt)
				node[anchor=north,font=\scriptsize] {$.4$};
				\draw (.6,0.3pt) -- (.6,-1pt)
				node[anchor=north,font=\scriptsize] {$.6$};
				\draw (.8,0.3pt) -- (.8,-1pt)
				node[anchor=north,font=\scriptsize] {$.8$};
				\draw (1,0.3pt) -- (1,-1pt)
				node[anchor=north,font=\scriptsize] {$1$};
				\draw (0.3pt,0) -- (-0.8pt,0);
				\draw (0.3pt,{1/5}) -- (-0.8pt,{1/5});
				\draw (0.3pt,{2/5}) -- (-0.8pt,{2/5});
				\draw (0.3pt,{3/5}) -- (-0.8pt,{3/5});
				\draw (0.3pt,{4/5}) -- (-0.8pt,{4/5});
				\draw (0.3pt,1) -- (-0.8pt,1);
				\node[left,font=\scriptsize] at (0,0) {$0$};
				\node[left,font=\scriptsize] at (0,{1/5}) {$.2$};
				\node[left,font=\scriptsize] at (0,{2/5}) {$.4$};
				\node[left,font=\scriptsize] at (0,{3/5}) {$.6$};
				\node[left,font=\scriptsize] at (0,{4/5}) {$.8$};
				\node[left,font=\scriptsize] at (0,1) {$1$};	
				\draw[red!100!blue,domain=0:1]  plot(\x, \x);	
				\draw[red!80!blue,domain=0:1]  plot(\x, {\x^1.2});	
				\draw[red!60!blue,domain=0:1]  plot(\x, {\x^1.4});	
				\draw[red!40!blue,domain=0:1]  plot(\x, {\x^1.6});	
				\draw[red!20!blue,domain=0:1]  plot(\x, {\x^1.8});	
				\draw[red!0!blue,domain=0:1]  plot(\x, {\x^2});
			\end{tikzpicture}	
			\begin{tikzpicture}[scale=3.5]
				\tikzstyle{vertex}=[font=\small,circle,draw,fill=yellow!20]
				\tikzstyle{edge} = [font=\scriptsize,draw,thick,-]
				\draw[black, thick] (0,0) -- (0,1);
				\draw[black, thick, ->] (0,0) -- (1.1,0);
				\draw (0,0.3pt) -- (0,-1pt)
				node[anchor=north,font=\scriptsize] {$0$};
				\draw (.2,0.3pt) -- (.2,-1pt)
				node[anchor=north,font=\scriptsize] {$.2$};
				\draw (.4,0.3pt) -- (.4,-1pt)
				node[anchor=north,font=\scriptsize] {$.4$};
				\draw (.6,0.3pt) -- (.6,-1pt)
				node[anchor=north,font=\scriptsize] {$.6$};
				\draw (.8,0.3pt) -- (.8,-1pt)
				node[anchor=north,font=\scriptsize] {$.8$};
				\draw (1,0.3pt) -- (1,-1pt)
				node[anchor=north,font=\scriptsize] {$1$};
				\draw (0.3pt,0) -- (-0.8pt,0);
				\draw (0.3pt,{1/4}) -- (-0.8pt,{1/4});
				\draw (0.3pt,{2/4}) -- (-0.8pt,{2/4});
				\draw (0.3pt,{3/4}) -- (-0.8pt,{3/4});
				\draw (0.3pt,1) -- (-0.8pt,1);
				\node[left,font=\scriptsize] at (0,0) {$0$};
				\node[left,font=\scriptsize] at (0,{1/4}) {$.025$};
				\node[left,font=\scriptsize] at (0,{2/4}) {$.05$};
				\node[left,font=\scriptsize] at (0,{3/4}) {$.075$};
				\node[left,font=\scriptsize] at (0,1) {$.1$};	
				\draw[red!0!blue] plot[smooth,mark=triangle,mark indices={1,2,3,6,9,12,15,18,21},mark size=1pt] file {size1bc6_smoothed.txt};	
				\draw[red!20!blue] plot[smooth] file {size1bc5_smoothed.txt};	
				\draw[red!40!blue] plot[smooth] file {size1bc4_smoothed.txt};	
				\draw[red!60!blue] plot[smooth] file {size1bc3_smoothed.txt};	
				\draw[red!80!blue] plot[smooth] file {size1bc2_smoothed.txt};	
				\draw[red!100!blue] plot[smooth] file {size1bc1_smoothed.txt};
				\node[right] at (1.1,0) {$\gamma$};
			\end{tikzpicture}	
			\begin{tikzpicture}[scale=3.5]
				\tikzstyle{vertex}=[font=\small,circle,draw,fill=yellow!20]
				\tikzstyle{edge} = [font=\scriptsize,draw,thick,-]
				\draw[black, thick] (0,0) -- (0,1);
				\draw[black, thick, ->] (0,0) -- (1.1,0);
				\draw (0,0.3pt) -- (0,-1pt)
				node[anchor=north,font=\scriptsize] {$0$};
				\draw (.2,0.3pt) -- (.2,-1pt)
				node[anchor=north,font=\scriptsize] {$.2$};
				\draw (.4,0.3pt) -- (.4,-1pt)
				node[anchor=north,font=\scriptsize] {$.4$};
				\draw (.6,0.3pt) -- (.6,-1pt)
				node[anchor=north,font=\scriptsize] {$.6$};
				\draw (.8,0.3pt) -- (.8,-1pt)
				node[anchor=north,font=\scriptsize] {$.8$};
				\draw (1,0.3pt) -- (1,-1pt)
				node[anchor=north,font=\scriptsize] {$1$};
				\draw (0.3pt,0) -- (-0.8pt,0);
				\draw (0.3pt,{1/4}) -- (-0.8pt,{1/4});
				\draw (0.3pt,{2/4}) -- (-0.8pt,{2/4});
				\draw (0.3pt,{3/4}) -- (-0.8pt,{3/4});
				\draw (0.3pt,1) -- (-0.8pt,1);
				\node[left,font=\scriptsize] at (0,0) {$0$};
				\node[left,font=\scriptsize] at (0,{1/4}) {$.025$};
				\node[left,font=\scriptsize] at (0,{2/4}) {$.05$};
				\node[left,font=\scriptsize] at (0,{3/4}) {$.075$};
				\node[left,font=\scriptsize] at (0,1) {$.1$};	
				\draw[red!100!blue] plot[smooth] file {size1bc2_smoothed.txt};	
				\draw[red!0!blue] plot[smooth,mark=asterisk,mark indices={1,2,3,6,9,12,15,18,21},mark size=1pt] file {size1dh_smoothed.txt};
				\node[right] at (1.1,0) {$\gamma$};
			\end{tikzpicture}	
			\begin{tikzpicture}[scale=3.5]
				\tikzstyle{vertex}=[font=\small,circle,draw,fill=yellow!20]
				\tikzstyle{edge} = [font=\scriptsize,draw,thick,-]
				\draw[black, thick] (0,0) -- (0,1);
				\draw[black, thick] (0,0) -- (1,0);
				\draw (0,0.3pt) -- (0,-1pt)
				node[anchor=north,font=\scriptsize] {$0$};
				\draw (.2,0.3pt) -- (.2,-1pt)
				node[anchor=north,font=\scriptsize] {$.2$};
				\draw (.4,0.3pt) -- (.4,-1pt)
				node[anchor=north,font=\scriptsize] {$.4$};
				\draw (.6,0.3pt) -- (.6,-1pt)
				node[anchor=north,font=\scriptsize] {$.6$};
				\draw (.8,0.3pt) -- (.8,-1pt)
				node[anchor=north,font=\scriptsize] {$.8$};
				\draw (1,0.3pt) -- (1,-1pt)
				node[anchor=north,font=\scriptsize] {$1$};
				\draw (0.3pt,0) -- (-0.8pt,0);
				\draw (0.3pt,{1/5}) -- (-0.8pt,{1/5});
				\draw (0.3pt,{2/5}) -- (-0.8pt,{2/5});
				\draw (0.3pt,{3/5}) -- (-0.8pt,{3/5});
				\draw (0.3pt,{4/5}) -- (-0.8pt,{4/5});
				\draw (0.3pt,1) -- (-0.8pt,1);
				\node[left,font=\scriptsize] at (0,0) {$0$};
				\node[left,font=\scriptsize] at (0,{1/5}) {$.2$};
				\node[left,font=\scriptsize] at (0,{2/5}) {$.4$};
				\node[left,font=\scriptsize] at (0,{3/5}) {$.6$};
				\node[left,font=\scriptsize] at (0,{4/5}) {$.8$};
				\node[left,font=\scriptsize] at (0,1) {$1$};	
				\draw[red!80!blue,domain=0:1] plot[smooth] file {ODC1.txt};
				\draw[red!60!blue,domain=0:1] plot[smooth] file {ODC2.txt};	
				\draw[red!40!blue,domain=0:1] plot[smooth] file {ODC3.txt};
				\draw[red!20!blue,domain=0:1] plot[smooth] file {ODC4.txt};
				\draw[red!0!blue,domain=0:1] plot[smooth] file {ODC5.txt};		
				\draw[red!100!blue,domain=0:1]  plot(\x, \x);
			\end{tikzpicture}	
			\begin{tikzpicture}[scale=3.5]
				\tikzstyle{vertex}=[font=\small,circle,draw,fill=yellow!20]
				\tikzstyle{edge} = [font=\scriptsize,draw,thick,-]
				\draw[black, thick] (0,0) -- (0,1);
				\draw[black, thick, ->] (0,0) -- (1.1,0);
				\draw (0,0.3pt) -- (0,-1pt)
				node[anchor=north,font=\scriptsize] {$0$};
				\draw (.2,0.3pt) -- (.2,-1pt)
				node[anchor=north,font=\scriptsize] {$.1$};
				\draw (.4,0.3pt) -- (.4,-1pt)
				node[anchor=north,font=\scriptsize] {$.2$};
				\draw (.6,0.3pt) -- (.6,-1pt)
				node[anchor=north,font=\scriptsize] {$.3$};
				\draw (.8,0.3pt) -- (.8,-1pt)
				node[anchor=north,font=\scriptsize] {$.4$};
				\draw (1,0.3pt) -- (1,-1pt)
				node[anchor=north,font=\scriptsize] {$.5$};
				\draw (0.3pt,0) -- (-0.8pt,0);
				\draw (0.3pt,{1/4}) -- (-0.8pt,{1/4});
				\draw (0.3pt,{2/4}) -- (-0.8pt,{2/4});
				\draw (0.3pt,{3/4}) -- (-0.8pt,{3/4});
				\draw (0.3pt,1) -- (-0.8pt,1);
				\node[left,font=\scriptsize] at (0,0) {$0$};
				\node[left,font=\scriptsize] at (0,{1/4}) {$.025$};
				\node[left,font=\scriptsize] at (0,{2/4}) {$.05$};
				\node[left,font=\scriptsize] at (0,{3/4}) {$.075$};
				\node[left,font=\scriptsize] at (0,1) {$.1$};
				\draw[red!0!blue] plot[smooth,mark=triangle,mark indices={1,4,7,10,13,16,19,22,25},mark size=1pt] file {size2bc6_smoothed.txt};	
				\draw[red!20!blue] plot[smooth] file {size2bc5_smoothed.txt};	
				\draw[red!40!blue] plot[smooth] file {size2bc4_smoothed.txt};	
				\draw[red!60!blue] plot[smooth] file {size2bc3_smoothed.txt};	
				\draw[red!80!blue] plot[smooth] file {size2bc2_smoothed.txt};	
				\draw[red!100!blue] plot[smooth] file {size2bc1_smoothed.txt};
				\node[right] at (1.1,0) {$\gamma$};
			\end{tikzpicture}	
			\begin{tikzpicture}[scale=3.5]
				\tikzstyle{vertex}=[font=\small,circle,draw,fill=yellow!20]
				\tikzstyle{edge} = [font=\scriptsize,draw,thick,-]
				\draw[black, thick] (0,0) -- (0,1);
				\draw[black, thick, ->] (0,0) -- (1.1,0);
				\draw (0,0.3pt) -- (0,-1pt)
				node[anchor=north,font=\scriptsize] {$0$};
				\draw (.2,0.3pt) -- (.2,-1pt)
				node[anchor=north,font=\scriptsize] {$.1$};
				\draw (.4,0.3pt) -- (.4,-1pt)
				node[anchor=north,font=\scriptsize] {$.2$};
				\draw (.6,0.3pt) -- (.6,-1pt)
				node[anchor=north,font=\scriptsize] {$.3$};
				\draw (.8,0.3pt) -- (.8,-1pt)
				node[anchor=north,font=\scriptsize] {$.4$};
				\draw (1,0.3pt) -- (1,-1pt)
				node[anchor=north,font=\scriptsize] {$.5$};
				\draw (0.3pt,0) -- (-0.8pt,0);
				\draw (0.3pt,{1/4}) -- (-0.8pt,{1/4});
				\draw (0.3pt,{2/4}) -- (-0.8pt,{2/4});
				\draw (0.3pt,{3/4}) -- (-0.8pt,{3/4});
				\draw (0.3pt,1) -- (-0.8pt,1);
				\node[left,font=\scriptsize] at (0,0) {$0$};
				\node[left,font=\scriptsize] at (0,{1/4}) {$.025$};
				\node[left,font=\scriptsize] at (0,{2/4}) {$.05$};
				\node[left,font=\scriptsize] at (0,{3/4}) {$.075$};
				\node[left,font=\scriptsize] at (0,1) {$.1$};		
				\draw[red!100!blue] plot[smooth] file {size2bc2_smoothed.txt};	
				\draw[red!0!blue] plot[smooth,mark=asterisk,mark indices={1,4,7,10,13,16,19,22,25},mark size=1pt] file {size2dh_smoothed.txt};
				\node[right] at (1.1,0) {$\gamma$};
			\end{tikzpicture}
			\caption{Null rejection frequencies using independent samples of size $n=500$ and nominal level $\alpha=.05$. P-P curves parametrized by $\gamma$, shifting away from the 45-degree line as $\gamma$ increases, are displayed in top-left and bottom-left panels. Rejection frequencies for the modified WMW test are displayed in top-center and bottom-center panels, with $\tau_n=.5,.75,1,1.25,1.5,\infty$, and triangles superimposed on the curves for $\tau_n=\infty$. Top-right and bottom-right panels show rejection frequencies for the modified WMW test with $\tau_n=.75$ and for the DH test with tuning parameter $-.135$, with asterisks superimposed on the curve for the latter test.}\label{fig:sizeinterior}
		\end{figure}
		
		In the top-center and bottom-center panels of \cref{fig:sizeinterior} we plot the rejection frequencies for the modified WMW test as a function of the parameter $\gamma$ for the P-P curves in the top-left and bottom-left panels. Separate curves are plotted for each of the tuning parameter values $\tau_n=.5,.75,1,1.25,1.5,\infty$, with the curves for smaller tuning parameter values lying above those for larger values. We superimpose triangles on the curve for $\tau_n=\infty$ to improve visibility. An interesting pattern is apparent wherein the rejection frequencies with $\tau_n<\infty$ initially decrease as we raise $\gamma$ above zero -- essentially decreasing to zero in the top-center panel -- before rising back toward the nominal level of .05 as $\gamma$ becomes larger. The rejection frequencies with $\tau_n=\infty$ decrease smoothly to zero in both the top-center and bottom-center panels. The results in the top-center panel are particularly encouraging because these null configurations do not belong to the boundary of the null and therefore, as discussed at the end of \cref{sec:modbsasy}, the modified bootstrap critical values are not guaranteed by \cref{prop:modtest} to produce limiting rejection frequencies no greater than the nominal level unless they are constrained not to fall below a small positive constant $\eta>0$. Similar to \citet{DH16}, we have found that placing a small positive lower bound such as $\eta=10^{-6}$ on the modified bootstrap critical values does not meaningfully affect rejection probabilities, so we simply set $\eta=0$.
		
		In the top-right and bottom-right panels of \cref{fig:sizeinterior} we plot the rejection frequencies obtained using the modified WMW test with $\tau_n=.75$ and the DH test with tuning parameter $-.139$ (the latter superimposed with asterisks) as a function of the parameter $\gamma$ for the P-P curves in the top-left and bottom-left panels. In the top-right panel we see that the rejection frequencies for the DH test do not share the interesting non-monotone behavior exhibited by the modified WMW test, and instead drop quickly to zero as we raise $\gamma$ above zero. On the other hand, we see in the bottom-right panel that the DH test is more successful than the WMW test in maintaining a rejection frequency close to the nominal level on the boundary of the null, at least for the family of P-P curves considered here.

		\subsubsection{Matched pairs sampling framework}
		\label{sec:numericalnullpairs}
		
		In \cref{lfctab_pairs} we report rejection frequencies at the least favorable case using the matched pairs sampling framework. These results may be compared directly to those reported in \cref{lfctab} for the independent sampling framework. The only difference between the simulation designs is that with matched pairs a Gaussian copula with parameter $\rho=.25,.5,.75$ is used to generate dependence within pairs and, as described in \cref{sec:modbs}, the empirical copula is used to estimate the contact set. The results in \cref{lfctab_pairs} are broadly similar to those in \cref{lfctab}, with the rejection frequencies using the modified WMW test tending to fall below the nominal level, and with the rejection frequencies using the DH test tending to fall modestly above the nominal level. The degree to which the rejection frequencies for the modified WMW test fall below the nominal level increases as the dependence between matched pairs increases.
		
		\begin{table}[!ht]
			\centering
			\begin{threeparttable}
				\caption{Rejection frequencies at the least favorable case with matched pairs.}\label{lfctab_pairs}
				{\small
					\begin{tabular}{crrcccccccccccc}
						\toprule
						\multicolumn{3}{c}{}&\multicolumn{6}{c}{Mod.\ Wilcoxon-Mann-Whitney}&\multicolumn{6}{c}{Donald-Hsu}\\
						\cmidrule(lr){4-9}\cmidrule(lr){10-15}
						\multicolumn{3}{r}{Tun.\ par.} & .5 & .75 & 1 & 1.25 & 1.5 & $\infty$ & -.025 & -.05 & -.1 & -.15 & -.2 & $-\infty$\\
						\midrule
						$\rho$&$\alpha$&$n$&&&&&&&&&&&\\
						\midrule
						\multirow{12}{*}{.25}&\multirow{6}{*}{.05}& 25 &5.2&4.3&3.8&3.5&3.3&3.2&7.6&7.6&7.6&7.6&7.5&7.1\\
						&& 50 &4.3&3.7&3.4&3.3&3.2&3.1&5.9&5.9&5.9&5.8&5.8&5.6\\
						&& 100 &4.7&4.2&3.9&3.8&3.8&3.7&5.7&5.7&5.7&5.6&5.6&5.3\\
						&& 200 &4.4&4.1&3.9&3.9&3.8&3.8&5.6&5.6&5.5&5.5&5.5&5.3\\
						&& 500 &5.0&4.7&4.5&4.4&4.3&4.3&5.5&5.5&5.5&5.4&5.4&5.2\\
						&& 1000 &5.0&4.6&4.5&4.4&4.3&4.3&5.1&5.1&5.0&5.0&5.0&4.8\\
						&\multirow{6}{*}{.01}& 25 &1.8&1.4&1.2&1.1&1.0&1.0&1.6&1.6&1.6&1.6&1.6&1.6\\
						&& 50 &1.1&0.9&0.8&0.8&0.8&0.8&1.3&1.3&1.3&1.3&1.3&1.3\\
						&& 100 &0.9&0.9&0.9&0.8&0.8&0.8&1.2&1.2&1.2&1.2&1.2&1.2\\
						&& 200 &0.9&0.8&0.8&0.8&0.8&0.8&1.1&1.1&1.1&1.1&1.1&1.1\\
						&& 500 &0.9&0.9&0.9&0.9&0.9&0.9&1.1&1.1&1.1&1.1&1.1&1.0\\
						&& 1000 &1.1&1.0&1.0&1.0&0.9&0.9&1.0&0.9&0.9&0.9&0.9&0.9\\
						\midrule
						\multirow{12}{*}{.5}&\multirow{6}{*}{.05}& 25 &4.6&3.6&3.0&2.9&2.6&2.4&7.2&7.2&7.2&7.2&7.0&6.6\\
						&& 50 &4.2&3.4&3.0&2.8&2.6&2.5&5.9&5.9&5.9&5.7&5.7&5.3\\
						&& 100 &4.9&4.2&3.9&3.6&3.5&3.4&6.4&6.4&6.4&6.2&6.2&5.7\\
						&& 200 &5.0&4.5&4.3&4.1&4.0&3.8&5.7&5.7&5.6&5.6&5.6&5.2\\
						&& 500 &5.4&4.9&4.5&4.3&4.2&4.1&5.4&5.4&5.4&5.3&5.3&5.1\\
						&& 1000 &6.1&5.5&5.2&5.0&4.9&4.8&6.1&6.0&6.0&5.9&5.9&5.6\\
						&\multirow{6}{*}{.01}& 25 &1.1&0.8&0.6&0.5&0.5&0.5&1.7&1.7&1.7&1.7&1.7&1.6\\
						&& 50 &0.8&0.6&0.6&0.5&0.5&0.5&1.2&1.2&1.2&1.1&1.1&1.1\\
						&& 100 &0.8&0.7&0.6&0.5&0.5&0.5&1.3&1.3&1.3&1.3&1.3&1.2\\
						&& 200 &0.9&0.8&0.7&0.7&0.7&0.7&1.2&1.2&1.2&1.2&1.2&1.2\\
						&& 500 &1.0&0.9&0.8&0.8&0.8&0.8&1.1&1.1&1.1&1.1&1.1&1.0\\
						&& 1000 &1.1&1.1&1.0&1.0&1.0&1.0&1.3&1.3&1.3&1.3&1.3&1.2\\
						\midrule
						\multirow{12}{*}{.75}&\multirow{6}{*}{.05}& 25 &2.9&1.8&1.4&1.3&1.1&0.9&6.1&6.1&6.1&6.1&5.9&5.1\\
						&& 50 &3.9&2.9&2.3&2.1&1.8&1.6&6.2&6.2&6.2&5.9&5.9&5.3\\
						&& 100 &4.5&3.3&2.9&2.6&2.4&2.2&6.3&6.3&6.2&6.0&5.9&5.3\\
						&& 200 &5.6&4.5&3.9&3.5&3.3&2.9&6.4&6.4&6.2&6.0&6.0&5.5\\
						&& 500 &5.3&4.2&3.8&3.5&3.3&3.1&5.9&5.9&5.8&5.7&5.6&5.0\\
						&& 1000 &6.0&5.0&4.6&4.3&4.1&3.9&5.7&5.7&5.5&5.5&5.4&5.0\\
						&\multirow{6}{*}{.01}& 25 &0.6&0.4&0.2&0.2&0.2&0.1&1.2&1.2&1.2&1.2&1.2&1.1\\
						&& 50 &0.4&0.3&0.2&0.2&0.2&0.1&1.0&1.0&1.0&0.9&0.9&0.8\\
						&& 100 &0.5&0.4&0.3&0.3&0.3&0.3&1.0&1.0&0.9&0.9&0.9&0.9\\
						&& 200 &0.8&0.7&0.6&0.6&0.5&0.5&1.1&1.1&1.1&1.1&1.1&0.9\\
						&& 500 &0.9&0.7&0.7&0.6&0.6&0.5&1.0&1.0&1.0&1.0&1.0&0.9\\
						&& 1000 &1.1&0.9&0.8&0.8&0.8&0.7&1.3&1.3&1.2&1.2&1.2&1.2\\
						\bottomrule
					\end{tabular}
				}
			\end{threeparttable}
		\end{table}
		
		In \cref{fig:sizeinterior_pairs} we report rejection frequencies with matched pairs ($n=500$, $\alpha=.05)$ corresponding to the two parametric families of P-P curves used to produce the results displayed in \cref{fig:sizeinterior}. The left, center and right columns of panels in \cref{fig:sizeinterior_pairs} correspond to the correlation parameter values $.25$, $.5$ and $.75$, respectively. The top (bottom) row of panels in \cref{fig:sizeinterior_pairs} corresponds to the family of P-P curves displayed in the top-left (bottom-left) panel in \cref{fig:sizeinterior}. Each panel of \cref{fig:sizeinterior_pairs} displays curves plotting the rejection frequency for three tests: the modified WMW test with $\tau_n=.75$ (superimposed with squares) and with $\tau_n=\infty$ (superimposed with triangles) and the DH test with tuning parameter $-.139$ (superimposed with asterisks). The results are very similar overall to those reported in \cref{fig:sizeinterior} for the independent sampling framework. The curves for the modified WMW test with $\tau_n=\infty$ and for the DH test are difficult to distinguish in the top row of panels.
		
		\begin{figure}
			\centering
			\begin{tikzpicture}[scale=3.5]
				\tikzstyle{vertex}=[font=\small,circle,draw,fill=yellow!20]
				\tikzstyle{edge} = [font=\scriptsize,draw,thick,-]
				\draw[black, thick] (0,0) -- (0,1);
				\draw[black, thick, ->] (0,0) -- (1.1,0);
				\draw (0,0.3pt) -- (0,-1pt)
				node[anchor=north,font=\scriptsize] {$0$};
				\draw (.2,0.3pt) -- (.2,-1pt)
				node[anchor=north,font=\scriptsize] {$.2$};
				\draw (.4,0.3pt) -- (.4,-1pt)
				node[anchor=north,font=\scriptsize] {$.4$};
				\draw (.6,0.3pt) -- (.6,-1pt)
				node[anchor=north,font=\scriptsize] {$.6$};
				\draw (.8,0.3pt) -- (.8,-1pt)
				node[anchor=north,font=\scriptsize] {$.8$};
				\draw (1,0.3pt) -- (1,-1pt)
				node[anchor=north,font=\scriptsize] {$1$};
				\draw (0.3pt,0) -- (-0.8pt,0);
				\draw (0.3pt,{1/4}) -- (-0.8pt,{1/4});
				\draw (0.3pt,{2/4}) -- (-0.8pt,{2/4});
				\draw (0.3pt,{3/4}) -- (-0.8pt,{3/4});
				\draw (0.3pt,1) -- (-0.8pt,1);
				\node[left,font=\scriptsize] at (0,0) {$0$};
				\node[left,font=\scriptsize] at (0,{1/4}) {$.025$};
				\node[left,font=\scriptsize] at (0,{2/4}) {$.05$};
				\node[left,font=\scriptsize] at (0,{3/4}) {$.075$};
				\node[left,font=\scriptsize] at (0,1) {$.1$};	
				\draw[red!0!blue] plot[smooth,mark=square,mark indices={6,9,12,15,18,21},mark size=.8pt] file {size1bc1_rho1_smoothed.txt};
				\draw[red!50!blue] plot[smooth,mark=triangle,mark indices={9,15,21},mark size=1pt] file {size1bc2_rho1_smoothed.txt};	
				\draw[red!100!blue] plot[smooth,mark=asterisk,mark indices={6,12,18},mark size=1pt] file {size1dh_rho1_smoothed.txt};
				\node[right] at (1.1,0) {$\gamma$};
			\end{tikzpicture}
			\begin{tikzpicture}[scale=3.5]
				\tikzstyle{vertex}=[font=\small,circle,draw,fill=yellow!20]
				\tikzstyle{edge} = [font=\scriptsize,draw,thick,-]
				\draw[black, thick] (0,0) -- (0,1);
				\draw[black, thick, ->] (0,0) -- (1.1,0);
				\draw (0,0.3pt) -- (0,-1pt)
				node[anchor=north,font=\scriptsize] {$0$};
				\draw (.2,0.3pt) -- (.2,-1pt)
				node[anchor=north,font=\scriptsize] {$.2$};
				\draw (.4,0.3pt) -- (.4,-1pt)
				node[anchor=north,font=\scriptsize] {$.4$};
				\draw (.6,0.3pt) -- (.6,-1pt)
				node[anchor=north,font=\scriptsize] {$.6$};
				\draw (.8,0.3pt) -- (.8,-1pt)
				node[anchor=north,font=\scriptsize] {$.8$};
				\draw (1,0.3pt) -- (1,-1pt)
				node[anchor=north,font=\scriptsize] {$1$};
				\draw (0.3pt,0) -- (-0.8pt,0);
				\draw (0.3pt,{1/4}) -- (-0.8pt,{1/4});
				\draw (0.3pt,{2/4}) -- (-0.8pt,{2/4});
				\draw (0.3pt,{3/4}) -- (-0.8pt,{3/4});
				\draw (0.3pt,1) -- (-0.8pt,1);
				\node[left,font=\scriptsize] at (0,0) {$0$};
				\node[left,font=\scriptsize] at (0,{1/4}) {$.025$};
				\node[left,font=\scriptsize] at (0,{2/4}) {$.05$};
				\node[left,font=\scriptsize] at (0,{3/4}) {$.075$};
				\node[left,font=\scriptsize] at (0,1) {$.1$};	
				\draw[red!0!blue] plot[smooth,mark=square,mark indices={6,9,12,15,18,21},mark size=.8pt] file {size1bc1_rho2_smoothed.txt};
				\draw[red!50!blue] plot[smooth,mark=triangle,mark indices={9,15,21},mark size=1pt] file {size1bc2_rho2_smoothed.txt};	
				\draw[red!100!blue] plot[smooth,mark=asterisk,mark indices={6,12,18},mark size=1pt] file {size1dh_rho2_smoothed.txt};
				\node[right] at (1.1,0) {$\gamma$};
			\end{tikzpicture}
			\begin{tikzpicture}[scale=3.5]
				\tikzstyle{vertex}=[font=\small,circle,draw,fill=yellow!20]
				\tikzstyle{edge} = [font=\scriptsize,draw,thick,-]
				\draw[black, thick] (0,0) -- (0,1);
				\draw[black, thick, ->] (0,0) -- (1.1,0);
				\draw (0,0.3pt) -- (0,-1pt)
				node[anchor=north,font=\scriptsize] {$0$};
				\draw (.2,0.3pt) -- (.2,-1pt)
				node[anchor=north,font=\scriptsize] {$.2$};
				\draw (.4,0.3pt) -- (.4,-1pt)
				node[anchor=north,font=\scriptsize] {$.4$};
				\draw (.6,0.3pt) -- (.6,-1pt)
				node[anchor=north,font=\scriptsize] {$.6$};
				\draw (.8,0.3pt) -- (.8,-1pt)
				node[anchor=north,font=\scriptsize] {$.8$};
				\draw (1,0.3pt) -- (1,-1pt)
				node[anchor=north,font=\scriptsize] {$1$};
				\draw (0.3pt,0) -- (-0.8pt,0);
				\draw (0.3pt,{1/4}) -- (-0.8pt,{1/4});
				\draw (0.3pt,{2/4}) -- (-0.8pt,{2/4});
				\draw (0.3pt,{3/4}) -- (-0.8pt,{3/4});
				\draw (0.3pt,1) -- (-0.8pt,1);
				\node[left,font=\scriptsize] at (0,0) {$0$};
				\node[left,font=\scriptsize] at (0,{1/4}) {$.025$};
				\node[left,font=\scriptsize] at (0,{2/4}) {$.05$};
				\node[left,font=\scriptsize] at (0,{3/4}) {$.075$};
				\node[left,font=\scriptsize] at (0,1) {$.1$};	
				\draw[red!0!blue] plot[smooth,mark=square,mark indices={6,9,12,15,18,21},mark size=.8pt] file {size1bc1_rho3_smoothed.txt};
				\draw[red!50!blue] plot[smooth,mark=triangle,mark indices={9,15,21},mark size=1pt] file {size1bc2_rho3_smoothed.txt};	
				\draw[red!100!blue] plot[smooth,mark=asterisk,mark indices={6,12,18},mark size=1pt] file {size1dh_rho3_smoothed.txt};
				\node[right] at (1.1,0) {$\gamma$};
			\end{tikzpicture}	
			\begin{tikzpicture}[scale=3.5]
				\tikzstyle{vertex}=[font=\small,circle,draw,fill=yellow!20]
				\tikzstyle{edge} = [font=\scriptsize,draw,thick,-]
				\draw[black, thick] (0,0) -- (0,1);
				\draw[black, thick, ->] (0,0) -- (1.1,0);
				\draw (0,0.3pt) -- (0,-1pt)
				node[anchor=north,font=\scriptsize] {$0$};
				\draw (.2,0.3pt) -- (.2,-1pt)
				node[anchor=north,font=\scriptsize] {$.1$};
				\draw (.4,0.3pt) -- (.4,-1pt)
				node[anchor=north,font=\scriptsize] {$.2$};
				\draw (.6,0.3pt) -- (.6,-1pt)
				node[anchor=north,font=\scriptsize] {$.3$};
				\draw (.8,0.3pt) -- (.8,-1pt)
				node[anchor=north,font=\scriptsize] {$.4$};
				\draw (1,0.3pt) -- (1,-1pt)
				node[anchor=north,font=\scriptsize] {$.5$};
				\draw (0.3pt,0) -- (-0.8pt,0);
				\draw (0.3pt,{1/4}) -- (-0.8pt,{1/4});
				\draw (0.3pt,{2/4}) -- (-0.8pt,{2/4});
				\draw (0.3pt,{3/4}) -- (-0.8pt,{3/4});
				\draw (0.3pt,1) -- (-0.8pt,1);
				\node[left,font=\scriptsize] at (0,0) {$0$};
				\node[left,font=\scriptsize] at (0,{1/4}) {$.025$};
				\node[left,font=\scriptsize] at (0,{2/4}) {$.05$};
				\node[left,font=\scriptsize] at (0,{3/4}) {$.075$};
				\node[left,font=\scriptsize] at (0,1) {$.1$};			
				\draw[red!0!blue] plot[smooth,mark=square,mark indices={4,7,10,13,16,19,22,25},mark size=.8pt] file {size2bc1_rho1_smoothed.txt};
				\draw[red!50!blue] plot[smooth,mark=triangle,mark indices={4,7,10,13,16,19,22,25},mark size=1pt] file {size2bc2_rho1_smoothed.txt};	
				\draw[red!100!blue] plot[smooth,mark=asterisk,mark indices={4,7,10,13,16,19,22,25},mark size=1pt] file {size2dh_rho1_smoothed.txt};
				\node[right] at (1.1,0) {$\gamma$};
			\end{tikzpicture}	
			\begin{tikzpicture}[scale=3.5]
				\tikzstyle{vertex}=[font=\small,circle,draw,fill=yellow!20]
				\tikzstyle{edge} = [font=\scriptsize,draw,thick,-]
				\draw[black, thick] (0,0) -- (0,1);
				\draw[black, thick, ->] (0,0) -- (1.1,0);
				\draw (0,0.3pt) -- (0,-1pt)
				node[anchor=north,font=\scriptsize] {$0$};
				\draw (.2,0.3pt) -- (.2,-1pt)
				node[anchor=north,font=\scriptsize] {$.1$};
				\draw (.4,0.3pt) -- (.4,-1pt)
				node[anchor=north,font=\scriptsize] {$.2$};
				\draw (.6,0.3pt) -- (.6,-1pt)
				node[anchor=north,font=\scriptsize] {$.3$};
				\draw (.8,0.3pt) -- (.8,-1pt)
				node[anchor=north,font=\scriptsize] {$.4$};
				\draw (1,0.3pt) -- (1,-1pt)
				node[anchor=north,font=\scriptsize] {$.5$};
				\draw (0.3pt,0) -- (-0.8pt,0);
				\draw (0.3pt,{1/4}) -- (-0.8pt,{1/4});
				\draw (0.3pt,{2/4}) -- (-0.8pt,{2/4});
				\draw (0.3pt,{3/4}) -- (-0.8pt,{3/4});
				\draw (0.3pt,1) -- (-0.8pt,1);
				\node[left,font=\scriptsize] at (0,0) {$0$};
				\node[left,font=\scriptsize] at (0,{1/4}) {$.025$};
				\node[left,font=\scriptsize] at (0,{2/4}) {$.05$};
				\node[left,font=\scriptsize] at (0,{3/4}) {$.075$};
				\node[left,font=\scriptsize] at (0,1) {$.1$};			
				\draw[red!0!blue] plot[smooth,mark=square,mark indices={4,7,10,13,16,19,22,25},mark size=.8pt] file {size2bc1_rho2_smoothed.txt};
				\draw[red!50!blue] plot[smooth,mark=triangle,mark indices={4,7,10,13,16,19,22,25},mark size=1pt] file {size2bc2_rho2_smoothed.txt};	
				\draw[red!100!blue] plot[smooth,mark=asterisk,mark indices={4,7,10,13,16,19,22,25},mark size=1pt] file {size2dh_rho2_smoothed.txt};
				\node[right] at (1.1,0) {$\gamma$};
			\end{tikzpicture}	
			\begin{tikzpicture}[scale=3.5]
				\tikzstyle{vertex}=[font=\small,circle,draw,fill=yellow!20]
				\tikzstyle{edge} = [font=\scriptsize,draw,thick,-]
				\draw[black, thick] (0,0) -- (0,1);
				\draw[black, thick, ->] (0,0) -- (1.1,0);
				\draw (0,0.3pt) -- (0,-1pt)
				node[anchor=north,font=\scriptsize] {$0$};
				\draw (.2,0.3pt) -- (.2,-1pt)
				node[anchor=north,font=\scriptsize] {$.1$};
				\draw (.4,0.3pt) -- (.4,-1pt)
				node[anchor=north,font=\scriptsize] {$.2$};
				\draw (.6,0.3pt) -- (.6,-1pt)
				node[anchor=north,font=\scriptsize] {$.3$};
				\draw (.8,0.3pt) -- (.8,-1pt)
				node[anchor=north,font=\scriptsize] {$.4$};
				\draw (1,0.3pt) -- (1,-1pt)
				node[anchor=north,font=\scriptsize] {$.5$};
				\draw (0.3pt,0) -- (-0.8pt,0);
				\draw (0.3pt,{1/4}) -- (-0.8pt,{1/4});
				\draw (0.3pt,{2/4}) -- (-0.8pt,{2/4});
				\draw (0.3pt,{3/4}) -- (-0.8pt,{3/4});
				\draw (0.3pt,1) -- (-0.8pt,1);
				\node[left,font=\scriptsize] at (0,0) {$0$};
				\node[left,font=\scriptsize] at (0,{1/4}) {$.025$};
				\node[left,font=\scriptsize] at (0,{2/4}) {$.05$};
				\node[left,font=\scriptsize] at (0,{3/4}) {$.075$};
				\node[left,font=\scriptsize] at (0,1) {$.1$};		
				\draw[red!0!blue] plot[smooth,mark=square,mark indices={4,7,10,13,16,19,22,25},mark size=.8pt] file {size2bc1_rho3_smoothed.txt};
				\draw[red!50!blue] plot[smooth,mark=triangle,mark indices={4,7,10,13,16,19,22,25},mark size=1pt] file {size2bc2_rho3_smoothed.txt};	
				\draw[red!100!blue] plot[smooth,mark=asterisk,mark indices={4,7,10,13,16,19,22,25},mark size=1pt] file {size2dh_rho3_smoothed.txt};
				\node[right] at (1.1,0) {$\gamma$};
			\end{tikzpicture}
			\caption{Null rejection frequencies using matched pair samples of size $n=500$ and nominal level $\alpha=.05$. The left, center and right columns of panels correspond to $\rho=.25,.5,.75$. Rejection frequencies are plotted for the modified WMW test with $\tau_n=.75$ (superimposed with squares) and with $\tau_n-\infty$ (superimposed with trangles), and the DH test with tuning parameter $-.139$ (superimposed with asterisks). The P-P curves, parametrized by $\gamma$, are displayed in top-left and bottom-left panels of \cref{fig:sizeinterior}.}\label{fig:sizeinterior_pairs}
		\end{figure}
				
		\subsection{Alternative rejection frequencies}
		\label{sec:numericalalt}
		
		\subsubsection{Independent sampling framework}
		\label{sec:numericalaltindep}
		
		In \cref{fig:power} we report rejection frequencies obtained with independent samples of size $n=500$ and nominal level $\alpha=.05$ for two parametric families of P-P curves not satisfying the null hypothesis of first-order stochastic dominance in general. The two families are displayed in the top-left and bottom-left panels of \cref{fig:power}. The family in the top-left has parametrization $R_\gamma(u)=u^{1-\gamma}$ with $\gamma\in[0,1)$, and the family in the bottom-left has parametrization $R_\gamma(u)=\Phi(\mathrm{e}^\gamma\Phi^{-1}(u))$ with $\gamma\in\mathbb R$. In both families we obtain the least favorable case when $\gamma=0$, while for other values of $\gamma$ the null hypothesis is not satisfied. The crucial difference between the two families is that, when the null hypothesis is not satisfied, the graph of $R_\gamma$ is everywhere above the 45-degree line (except at the endpoints zero and one) with the top-left family but is partially below the 45-degree line with the bottom-left family.
				
		\begin{figure}[!ht]
			\centering
				\begin{tikzpicture}[scale=3.5]
					\tikzstyle{vertex}=[font=\small,circle,draw,fill=yellow!20]
					\tikzstyle{edge} = [font=\scriptsize,draw,thick,-]
					\draw[black, thick] (0,0) -- (0,1);
					\draw[black, thick] (0,0) -- (1,0);
					\draw (0,0.3pt) -- (0,-1pt)
					node[anchor=north,font=\scriptsize] {$0$};
					\draw (.2,0.3pt) -- (.2,-1pt)
					node[anchor=north,font=\scriptsize] {$.2$};
					\draw (.4,0.3pt) -- (.4,-1pt)
					node[anchor=north,font=\scriptsize] {$.4$};
					\draw (.6,0.3pt) -- (.6,-1pt)
					node[anchor=north,font=\scriptsize] {$.6$};
					\draw (.8,0.3pt) -- (.8,-1pt)
					node[anchor=north,font=\scriptsize] {$.8$};
					\draw (1,0.3pt) -- (1,-1pt)
					node[anchor=north,font=\scriptsize] {$1$};
					\draw (0.3pt,0) -- (-0.8pt,0);
					\draw (0.3pt,{1/5}) -- (-0.8pt,{1/5});
					\draw (0.3pt,{2/5}) -- (-0.8pt,{2/5});
					\draw (0.3pt,{3/5}) -- (-0.8pt,{3/5});
					\draw (0.3pt,{4/5}) -- (-0.8pt,{4/5});
					\draw (0.3pt,1) -- (-0.8pt,1);
					\node[left,font=\scriptsize] at (0,0) {$0$};
					\node[left,font=\scriptsize] at (0,{1/5}) {$.2$};
					\node[left,font=\scriptsize] at (0,{2/5}) {$.4$};
					\node[left,font=\scriptsize] at (0,{3/5}) {$.6$};
					\node[left,font=\scriptsize] at (0,{4/5}) {$.8$};
					\node[left,font=\scriptsize] at (0,1) {$1$};	
					\draw[red!0!blue,domain=0:1]  plot(\x, \x);	
					\draw[red!20!blue,domain=0:1]  plot(\x, {\x^(1-.05)});	
					\draw[red!40!blue,domain=0:1]  plot(\x, {\x^(1-.1)});	
					\draw[red!60!blue,domain=0:1]  plot(\x, {\x^(1-.15)});	
					\draw[red!80!blue,domain=0:1]  plot(\x, {\x^(1-.2)});	
					\draw[red!100!blue,domain=0:1]  plot(\x, {\x^(1-.25)});
				\end{tikzpicture}	
				\begin{tikzpicture}[scale=3.5]
					\tikzstyle{vertex}=[font=\small,circle,draw,fill=yellow!20]
					\tikzstyle{edge} = [font=\scriptsize,draw,thick,-]
					\draw[black, thick] (0,0) -- (0,1);
					\draw[black, thick, ->] (0,0) -- (1.1,0);
					\draw (0,0.3pt) -- (0,-1pt)
					node[anchor=north,font=\scriptsize] {$0$};
					\draw (.2,0.3pt) -- (.2,-1pt)
					node[anchor=north,font=\scriptsize] {$.05$};
					\draw (.4,0.3pt) -- (.4,-1pt)
					node[anchor=north,font=\scriptsize] {$.1$};
					\draw (.6,0.3pt) -- (.6,-1pt)
					node[anchor=north,font=\scriptsize] {$.15$};
					\draw (.8,0.3pt) -- (.8,-1pt)
					node[anchor=north,font=\scriptsize] {$.2$};
					\draw (1,0.3pt) -- (1,-1pt)
					node[anchor=north,font=\scriptsize] {$.25$};
					\draw (0.3pt,0) -- (-0.8pt,0);
					\draw (0.3pt,{1/5}) -- (-0.8pt,{1/5});
					\draw (0.3pt,{2/5}) -- (-0.8pt,{2/5});
					\draw (0.3pt,{3/5}) -- (-0.8pt,{3/5});
					\draw (0.3pt,{4/5}) -- (-0.8pt,{4/5});
					\draw (0.3pt,1) -- (-0.8pt,1);
					\node[left,font=\scriptsize] at (0,0) {$0$};
					\node[left,font=\scriptsize] at (0,{1/5}) {$.2$};
					\node[left,font=\scriptsize] at (0,{2/5}) {$.4$};
					\node[left,font=\scriptsize] at (0,{3/5}) {$.6$};
					\node[left,font=\scriptsize] at (0,{4/5}) {$.8$};
					\node[left,font=\scriptsize] at (0,1) {$1$};	
					\draw[red!0!blue] plot[smooth,mark=triangle,mark indices={2,5,8,11,14,17,20,23,26},mark size=1pt] file {power1bc6_rev.txt};	
					\draw[red!20!blue] plot[smooth] file {power1bc5_rev.txt};	
					\draw[red!40!blue] plot[smooth] file {power1bc4_rev.txt};	
					\draw[red!60!blue] plot[smooth] file {power1bc3_rev.txt};	
					\draw[red!80!blue] plot[smooth] file {power1bc2_rev.txt};	
					\draw[red!100!blue] plot[smooth] file {power1bc1_rev.txt};
					\node[right] at (1.1,0) {$\gamma$};
				\end{tikzpicture}	
				\begin{tikzpicture}[scale=3.5]
					\tikzstyle{vertex}=[font=\small,circle,draw,fill=yellow!20]
					\tikzstyle{edge} = [font=\scriptsize,draw,thick,-]
					\draw[black, thick] (0,0) -- (0,1);
					\draw[black, thick, ->] (0,0) -- (1.1,0);
					\draw (0,0.3pt) -- (0,-1pt)
					node[anchor=north,font=\scriptsize] {$0$};
					\draw (.2,0.3pt) -- (.2,-1pt)
					node[anchor=north,font=\scriptsize] {$.05$};
					\draw (.4,0.3pt) -- (.4,-1pt)
					node[anchor=north,font=\scriptsize] {$.1$};
					\draw (.6,0.3pt) -- (.6,-1pt)
					node[anchor=north,font=\scriptsize] {$.15$};
					\draw (.8,0.3pt) -- (.8,-1pt)
					node[anchor=north,font=\scriptsize] {$.2$};
					\draw (1,0.3pt) -- (1,-1pt)
					node[anchor=north,font=\scriptsize] {$.25$};
					\draw (0.3pt,0) -- (-0.8pt,0);
					\draw (0.3pt,{1/5}) -- (-0.8pt,{1/5});
					\draw (0.3pt,{2/5}) -- (-0.8pt,{2/5});
					\draw (0.3pt,{3/5}) -- (-0.8pt,{3/5});
					\draw (0.3pt,{4/5}) -- (-0.8pt,{4/5});
					\draw (0.3pt,1) -- (-0.8pt,1);
					\node[left,font=\scriptsize] at (0,0) {$0$};
					\node[left,font=\scriptsize] at (0,{1/5}) {$.2$};
					\node[left,font=\scriptsize] at (0,{2/5}) {$.4$};
					\node[left,font=\scriptsize] at (0,{3/5}) {$.6$};
					\node[left,font=\scriptsize] at (0,{4/5}) {$.8$};
					\node[left,font=\scriptsize] at (0,1) {$1$};		
					\draw[red!0!blue] plot[smooth] file {power1bc2_rev.txt};	
					\draw[red!100!blue] plot[smooth,mark=asterisk,mark indices={2,5,8,11,14,17,20,23,26},mark size=1pt] file {power1dh_rev.txt};
					\node[right] at (1.1,0) {$\gamma$};
				\end{tikzpicture}	
				\begin{tikzpicture}[scale=3.5]
					\tikzstyle{vertex}=[font=\small,circle,draw,fill=yellow!20]
					\tikzstyle{edge} = [font=\scriptsize,draw,thick,-]
					\draw[black, thick] (0,0) -- (0,1);
					\draw[black, thick] (0,0) -- (1,0);
					\draw (0,0.3pt) -- (0,-1pt)
					node[anchor=north,font=\scriptsize] {$0$};
					\draw (.2,0.3pt) -- (.2,-1pt)
					node[anchor=north,font=\scriptsize] {$.2$};
					\draw (.4,0.3pt) -- (.4,-1pt)
					node[anchor=north,font=\scriptsize] {$.4$};
					\draw (.6,0.3pt) -- (.6,-1pt)
					node[anchor=north,font=\scriptsize] {$.6$};
					\draw (.8,0.3pt) -- (.8,-1pt)
					node[anchor=north,font=\scriptsize] {$.8$};
					\draw (1,0.3pt) -- (1,-1pt)
					node[anchor=north,font=\scriptsize] {$1$};
					\draw (0.3pt,0) -- (-0.8pt,0);
					\draw (0.3pt,{1/5}) -- (-0.8pt,{1/5});
					\draw (0.3pt,{2/5}) -- (-0.8pt,{2/5});
					\draw (0.3pt,{3/5}) -- (-0.8pt,{3/5});
					\draw (0.3pt,{4/5}) -- (-0.8pt,{4/5});
					\draw (0.3pt,1) -- (-0.8pt,1);
					\node[left,font=\scriptsize] at (0,0) {$0$};
					\node[left,font=\scriptsize] at (0,{1/5}) {$.2$};
					\node[left,font=\scriptsize] at (0,{2/5}) {$.4$};
					\node[left,font=\scriptsize] at (0,{3/5}) {$.6$};
					\node[left,font=\scriptsize] at (0,{4/5}) {$.8$};
					\node[left,font=\scriptsize] at (0,1) {$1$};	
					\draw[red!0!blue,domain=0:1]  plot(\x, \x);
					\draw[red!20!blue,domain=0:1] plot[smooth] file {ODC6.txt};
					\draw[red!40!blue,domain=0:1] plot[smooth] file {ODC7.txt};	
					\draw[red!60!blue,domain=0:1] plot[smooth] file {ODC8.txt};
					\draw[red!80!blue,domain=0:1] plot[smooth] file {ODC9.txt};
					\draw[red!100!blue,domain=0:1] plot[smooth] file {ODC10.txt};
					\draw[red!20!blue,dotted,domain=0:1] plot[smooth] file {ODC11.txt};
					\draw[red!40!blue,dotted,domain=0:1] plot[smooth] file {ODC12.txt};	
					\draw[red!60!blue,dotted,domain=0:1] plot[smooth] file {ODC13.txt};
					\draw[red!80!blue,dotted,domain=0:1] plot[smooth] file {ODC14.txt};
					\draw[red!100!blue,dotted,domain=0:1] plot[smooth] file {ODC15.txt};
				\end{tikzpicture}	
				\begin{tikzpicture}[scale=3.5]
					\tikzstyle{vertex}=[font=\small,circle,draw,fill=yellow!20]
					\tikzstyle{edge} = [font=\scriptsize,draw,thick,-]
					\draw[black, thick] (0,0) -- (0,1);
					\draw[black, thick, ->] (0,0) -- (1.1,0);
					\draw (0,0.3pt) -- (0,-1pt)
					node[anchor=north,font=\scriptsize] {$-.5$};
					\draw (.25,0.3pt) -- (.25,-1pt)
					node[anchor=north,font=\scriptsize] {$-.25$};
					\draw (.5,0.3pt) -- (.5,-1pt)
					node[anchor=north,font=\scriptsize] {$0$};
					\draw (.75,0.3pt) -- (.75,-1pt)
					node[anchor=north,font=\scriptsize] {$.25$};
					\draw (1,0.3pt) -- (1,-1pt)
					node[anchor=north,font=\scriptsize] {$.5$};
					\draw (0.3pt,0) -- (-0.8pt,0);
					\draw (0.3pt,{1/5}) -- (-0.8pt,{1/5});
					\draw (0.3pt,{2/5}) -- (-0.8pt,{2/5});
					\draw (0.3pt,{3/5}) -- (-0.8pt,{3/5});
					\draw (0.3pt,{4/5}) -- (-0.8pt,{4/5});
					\draw (0.3pt,1) -- (-0.8pt,1);
					\node[left,font=\scriptsize] at (0,0) {$0$};
					\node[left,font=\scriptsize] at (0,{1/5}) {$.2$};
					\node[left,font=\scriptsize] at (0,{2/5}) {$.4$};
					\node[left,font=\scriptsize] at (0,{3/5}) {$.6$};
					\node[left,font=\scriptsize] at (0,{4/5}) {$.8$};
					\node[left,font=\scriptsize] at (0,1) {$1$};	
					\draw[red!0!blue] plot[smooth,mark=triangle,mark indices={1,3,5,7,9,11,13,15,17,19,21},mark size=1pt] file {power2bc6.txt};	
					\draw[red!20!blue] plot[smooth] file {power2bc5.txt};	
					\draw[red!40!blue] plot[smooth] file {power2bc4.txt};	
					\draw[red!60!blue] plot[smooth] file {power2bc3.txt};	
					\draw[red!80!blue] plot[smooth] file {power2bc2.txt};	
					\draw[red!100!blue] plot[smooth] file {power2bc1.txt};
					\node[right] at (1.1,0) {$\gamma$};
				\end{tikzpicture}	
				\begin{tikzpicture}[scale=3.5]
					\tikzstyle{vertex}=[font=\small,circle,draw,fill=yellow!20]
					\tikzstyle{edge} = [font=\scriptsize,draw,thick,-]
					\draw[black, thick] (0,0) -- (0,1);
					\draw[black, thick, ->] (0,0) -- (1.1,0);
					\draw (0,0.3pt) -- (0,-1pt)
					node[anchor=north,font=\scriptsize] {$-.5$};
					\draw (.25,0.3pt) -- (.25,-1pt)
					node[anchor=north,font=\scriptsize] {$-.25$};
					\draw (.5,0.3pt) -- (.5,-1pt)
					node[anchor=north,font=\scriptsize] {$0$};
					\draw (.75,0.3pt) -- (.75,-1pt)
					node[anchor=north,font=\scriptsize] {$.25$};
					\draw (1,0.3pt) -- (1,-1pt)
					node[anchor=north,font=\scriptsize] {$.5$};
					\draw (0.3pt,0) -- (-0.8pt,0);
					\draw (0.3pt,{1/5}) -- (-0.8pt,{1/5});
					\draw (0.3pt,{2/5}) -- (-0.8pt,{2/5});
					\draw (0.3pt,{3/5}) -- (-0.8pt,{3/5});
					\draw (0.3pt,{4/5}) -- (-0.8pt,{4/5});
					\draw (0.3pt,1) -- (-0.8pt,1);
					\node[left,font=\scriptsize] at (0,0) {$0$};
					\node[left,font=\scriptsize] at (0,{1/5}) {$.2$};
					\node[left,font=\scriptsize] at (0,{2/5}) {$.4$};
					\node[left,font=\scriptsize] at (0,{3/5}) {$.6$};
					\node[left,font=\scriptsize] at (0,{4/5}) {$.8$};
					\node[left,font=\scriptsize] at (0,1) {$1$};		
					\draw[red!0!blue] plot[smooth] file {power2bc2.txt};	
					\draw[red!100!blue] plot[smooth,mark=asterisk,mark indices={1,3,5,7,9,11,13,15,17,19,21},mark size=1pt] file {power2dh.txt};
					\node[right] at (1.1,0) {$\gamma$};
				\end{tikzpicture}
			\caption{Alternative rejection frequencies using independent samples of size $n=500$ and nominal level $\alpha=.05$. P-P curves parametrized by $\gamma$, shifting away from the 45-degree line as $\gamma$ increases in magnitude, are displayed in top-left and bottom-left panels. Rejection frequencies for the modified WMW test are displayed in top-center and bottom-center panels, with $\tau_n=.5,.75,1,1.25,1.5,\infty$, and triangles superimposed on the curves for $\tau_n=\infty$. Top-right and bottom-right panels show rejection frequencies for the modified WMW test with $\tau_n=.75$ and for the DH test with tuning parameter $-.139$, with asterisks superimposed on the curves for the DH test.}\label{fig:power}
		\end{figure}
				
		The top-center and bottom-center panels in \cref{fig:power} display the rejection frequencies for the modified WMW test with tuning parameter values $\tau_n=.5,.75,1.1.25,1.5,\infty$. The curve for $\tau_n=\infty$, which corresponds to the standard bootstrap critical value, has triangles superimposed. In both panels we see the rejection frequencies increase from approximately $\alpha=.05$ to one as $\gamma$ moves away from zero, reflecting the consistency of the tests. In the top-center panel the curves plotted for different tuning parameter values are indistinguishable. In the bottom-center panel there is a clear separation between the curves, with the rejection frequencies using the standard bootstrap critical value well below the rejection frequencies using the modified bootstrap critical value. The very different behavior displayed in the two panels can be understood by observing that the modified bootstrap statistic $\tilde{S}^\ast$ differs from the standard bootstrap statistic $\hat{S}^\ast$ only when the P-P plot $\hat{R}$ falls below the 45-degree line by more than some threshold depending on the tuning parameter. This happens very infrequently in the simulations generating the rejection frequencies in the top-center panel because here we have $R_\gamma(u)>u$ for all $u\in(0,1)$ when $\gamma>0$; but frequently in those generating the rejection frequencies in the bottom-center panel because here we have $R_\gamma(u)<u$ for all $u\in(.5,1)$ when $\gamma<0$, and $R_\gamma(u)<u$ for all $u\in(0,.5)$ when $\gamma>0$.
		
		In the top-right and bottom-right panels of \cref{fig:power} we plot the rejection frequencies for the modified WMW test with $\tau_n=.75$ and the DH test with tuning parameter $-.139$ against one another. Rejection frequencies for the DH test are superimposed with asterisks. We see that the rejection frequencies for the two tests are similar. A slight power advantage for the modified WMW test is observed in the top-right panel, and a slight power advantage for the DH test in the bottom-right panel.
		
		\subsubsection{Matched pairs sampling framework}
		\label{sec:numericalaltpairs}
		
		\cref{fig:power_pairs} shows how the rejection frequencies plotted in \cref{fig:power} are affected when there is positive dependence between paired observations. The left, center and right columns of panels in \cref{fig:power_pairs} correspond to $\rho=.25,.5,.75$ respectively, while the top (bottom) row of panels corresponds to the family of P-P curves displayed in the top-left (bottom-left) panel in \cref{fig:power}. Rejection frequencies are plotted for the modified WMW test with $\tau_n=.75$ (superimposed with squares) and with $\tau_n=\infty$ (superimposed with triangles), and for the DH test with tuning parameter $-.139$ (superimposed with asterisks). In both rows of panels we see that the power for all three tests improves as $\rho$ increases. In the top row of panels we see that the rejection rates for the modified WMW test with $\tau_n=.75$ and with $\tau_n=\infty$ are indistinguishable, as they were with independent samples in \cref{fig:power}. The slight power advantage of these tests over the DH test widens as $\rho$ increases, becoming quite substantial with $\rho=.75$. In the bottom row of panels we see that the slight power advantage of the DH test observed with independent samples erodes as the correlation parameter increases. The rejection frequencies for the modified WMW test with $\tau_n=.75$ and the DH test are nearly indistinguishable when $\rho=.5$ or $\rho=.75$.
		
		\begin{figure}
			\centering
			\begin{tikzpicture}[scale=3.5]
				\tikzstyle{vertex}=[font=\small,circle,draw,fill=yellow!20]
				\tikzstyle{edge} = [font=\scriptsize,draw,thick,-]
				\draw[black, thick] (0,0) -- (0,1);
				\draw[black, thick, ->] (0,0) -- (1.1,0);
				\draw (0,0.3pt) -- (0,-1pt)
				node[anchor=north,font=\scriptsize] {$0$};
				\draw (.2,0.3pt) -- (.2,-1pt)
				node[anchor=north,font=\scriptsize] {$.05$};
				\draw (.4,0.3pt) -- (.4,-1pt)
				node[anchor=north,font=\scriptsize] {$.1$};
				\draw (.6,0.3pt) -- (.6,-1pt)
				node[anchor=north,font=\scriptsize] {$.15$};
				\draw (.8,0.3pt) -- (.8,-1pt)
				node[anchor=north,font=\scriptsize] {$.2$};
				\draw (1,0.3pt) -- (1,-1pt)
				node[anchor=north,font=\scriptsize] {$.25$};
				\draw (0.3pt,0) -- (-0.8pt,0);
				\draw (0.3pt,{1/5}) -- (-0.8pt,{1/5});
				\draw (0.3pt,{2/5}) -- (-0.8pt,{2/5});
				\draw (0.3pt,{3/5}) -- (-0.8pt,{3/5});
				\draw (0.3pt,{4/5}) -- (-0.8pt,{4/5});
				\draw (0.3pt,1) -- (-0.8pt,1);
				\node[left,font=\scriptsize] at (0,0) {$0$};
				\node[left,font=\scriptsize] at (0,{1/5}) {$.2$};
				\node[left,font=\scriptsize] at (0,{2/5}) {$.4$};
				\node[left,font=\scriptsize] at (0,{3/5}) {$.6$};
				\node[left,font=\scriptsize] at (0,{4/5}) {$.8$};
				\node[left,font=\scriptsize] at (0,1) {$1$};		
				\draw[red!0!blue] plot[smooth,mark=square,mark indices={9,15},mark size=.8pt] file {power1bc1_rho1_rev.txt};	
				\draw[red!50!blue] plot[smooth,mark=triangle,mark indices={12,19},mark size=1pt] file {power1bc2_rho1_rev.txt};	
				\draw[red!100!blue] plot[smooth,mark=asterisk,mark indices={7,11,14,17,21},mark size=1pt] file {power1dh_rho1_rev.txt};
				\node[right] at (1.1,0) {$\gamma$};
			\end{tikzpicture}	
			\begin{tikzpicture}[scale=3.5]
				\tikzstyle{vertex}=[font=\small,circle,draw,fill=yellow!20]
				\tikzstyle{edge} = [font=\scriptsize,draw,thick,-]
				\draw[black, thick] (0,0) -- (0,1);
				\draw[black, thick, ->] (0,0) -- (1.1,0);
				\draw (0,0.3pt) -- (0,-1pt)
				node[anchor=north,font=\scriptsize] {$0$};
				\draw (.2,0.3pt) -- (.2,-1pt)
				node[anchor=north,font=\scriptsize] {$.05$};
				\draw (.4,0.3pt) -- (.4,-1pt)
				node[anchor=north,font=\scriptsize] {$.1$};
				\draw (.6,0.3pt) -- (.6,-1pt)
				node[anchor=north,font=\scriptsize] {$.15$};
				\draw (.8,0.3pt) -- (.8,-1pt)
				node[anchor=north,font=\scriptsize] {$.2$};
				\draw (1,0.3pt) -- (1,-1pt)
				node[anchor=north,font=\scriptsize] {$.25$};
				\draw (0.3pt,0) -- (-0.8pt,0);
				\draw (0.3pt,{1/5}) -- (-0.8pt,{1/5});
				\draw (0.3pt,{2/5}) -- (-0.8pt,{2/5});
				\draw (0.3pt,{3/5}) -- (-0.8pt,{3/5});
				\draw (0.3pt,{4/5}) -- (-0.8pt,{4/5});
				\draw (0.3pt,1) -- (-0.8pt,1);
				\node[left,font=\scriptsize] at (0,0) {$0$};
				\node[left,font=\scriptsize] at (0,{1/5}) {$.2$};
				\node[left,font=\scriptsize] at (0,{2/5}) {$.4$};
				\node[left,font=\scriptsize] at (0,{3/5}) {$.6$};
				\node[left,font=\scriptsize] at (0,{4/5}) {$.8$};
				\node[left,font=\scriptsize] at (0,1) {$1$};		
				\draw[red!0!blue] plot[smooth,mark=square,mark indices={9,15},mark size=.8pt] file {power1bc1_rho2_rev.txt};	
				\draw[red!50!blue] plot[smooth,mark=triangle,mark indices={6,12,19},mark size=1pt] file {power1bc2_rho2_rev.txt};	
				\draw[red!100!blue] plot[smooth,mark=asterisk,mark indices={8,11,14,17},mark size=1pt] file {power1dh_rho2_rev.txt};
				\node[right] at (1.1,0) {$\gamma$};
			\end{tikzpicture}	
			\begin{tikzpicture}[scale=3.5]
				\tikzstyle{vertex}=[font=\small,circle,draw,fill=yellow!20]
				\tikzstyle{edge} = [font=\scriptsize,draw,thick,-]
				\draw[black, thick] (0,0) -- (0,1);
				\draw[black, thick, ->] (0,0) -- (1.1,0);
				\draw (0,0.3pt) -- (0,-1pt)
				node[anchor=north,font=\scriptsize] {$0$};
				\draw (.2,0.3pt) -- (.2,-1pt)
				node[anchor=north,font=\scriptsize] {$.05$};
				\draw (.4,0.3pt) -- (.4,-1pt)
				node[anchor=north,font=\scriptsize] {$.1$};
				\draw (.6,0.3pt) -- (.6,-1pt)
				node[anchor=north,font=\scriptsize] {$.15$};
				\draw (.8,0.3pt) -- (.8,-1pt)
				node[anchor=north,font=\scriptsize] {$.2$};
				\draw (1,0.3pt) -- (1,-1pt)
				node[anchor=north,font=\scriptsize] {$.25$};
				\draw (0.3pt,0) -- (-0.8pt,0);
				\draw (0.3pt,{1/5}) -- (-0.8pt,{1/5});
				\draw (0.3pt,{2/5}) -- (-0.8pt,{2/5});
				\draw (0.3pt,{3/5}) -- (-0.8pt,{3/5});
				\draw (0.3pt,{4/5}) -- (-0.8pt,{4/5});
				\draw (0.3pt,1) -- (-0.8pt,1);
				\node[left,font=\scriptsize] at (0,0) {$0$};
				\node[left,font=\scriptsize] at (0,{1/5}) {$.2$};
				\node[left,font=\scriptsize] at (0,{2/5}) {$.4$};
				\node[left,font=\scriptsize] at (0,{3/5}) {$.6$};
				\node[left,font=\scriptsize] at (0,{4/5}) {$.8$};
				\node[left,font=\scriptsize] at (0,1) {$1$};		
				\draw[red!0!blue] plot[smooth,mark=square,mark indices={8,14},mark size=.8pt] file {power1bc1_rho3_rev.txt};	
				\draw[red!50!blue] plot[smooth,mark=triangle,mark indices={6,11},mark size=1pt] file {power1bc2_rho3_rev.txt};	
				\draw[red!100!blue] plot[smooth,mark=asterisk,mark indices={8,11,14},mark size=1pt] file {power1dh_rho3_rev.txt};
				\node[right] at (1.1,0) {$\gamma$};
			\end{tikzpicture}	
			\begin{tikzpicture}[scale=3.5]
				\tikzstyle{vertex}=[font=\small,circle,draw,fill=yellow!20]
				\tikzstyle{edge} = [font=\scriptsize,draw,thick,-]
				\draw[black, thick] (0,0) -- (0,1);
				\draw[black, thick, ->] (0,0) -- (1.1,0);
				\draw (0,0.3pt) -- (0,-1pt)
				node[anchor=north,font=\scriptsize] {$-.5$};
				\draw (.25,0.3pt) -- (.25,-1pt)
				node[anchor=north,font=\scriptsize] {$-.25$};
				\draw (.5,0.3pt) -- (.5,-1pt)
				node[anchor=north,font=\scriptsize] {$0$};
				\draw (.75,0.3pt) -- (.75,-1pt)
				node[anchor=north,font=\scriptsize] {$.25$};
				\draw (1,0.3pt) -- (1,-1pt)
				node[anchor=north,font=\scriptsize] {$.5$};
				\draw (0.3pt,0) -- (-0.8pt,0);
				\draw (0.3pt,{1/5}) -- (-0.8pt,{1/5});
				\draw (0.3pt,{2/5}) -- (-0.8pt,{2/5});
				\draw (0.3pt,{3/5}) -- (-0.8pt,{3/5});
				\draw (0.3pt,{4/5}) -- (-0.8pt,{4/5});
				\draw (0.3pt,1) -- (-0.8pt,1);
				\node[left,font=\scriptsize] at (0,0) {$0$};
				\node[left,font=\scriptsize] at (0,{1/5}) {$.2$};
				\node[left,font=\scriptsize] at (0,{2/5}) {$.4$};
				\node[left,font=\scriptsize] at (0,{3/5}) {$.6$};
				\node[left,font=\scriptsize] at (0,{4/5}) {$.8$};
				\node[left,font=\scriptsize] at (0,1) {$1$};		
				\draw[red!0!blue] plot[smooth,mark=square,mark indices={3,7,14,16},mark size=.8pt] file {power2bc1_rho1.txt};
				\draw[red!50!blue] plot[smooth,mark=triangle,mark indices={2,4,6,16,18,20},mark size=1pt] file {power2bc2_rho1.txt};	
				\draw[red!100!blue] plot[smooth,mark=asterisk,mark indices={6,8,15,19},mark size=1pt] file {power2dh_rho1.txt};
				\node[right] at (1.1,0) {$\gamma$};
			\end{tikzpicture}
			\begin{tikzpicture}[scale=3.5]
				\tikzstyle{vertex}=[font=\small,circle,draw,fill=yellow!20]
				\tikzstyle{edge} = [font=\scriptsize,draw,thick,-]
				\draw[black, thick] (0,0) -- (0,1);
				\draw[black, thick, ->] (0,0) -- (1.1,0);
				\draw (0,0.3pt) -- (0,-1pt)
				node[anchor=north,font=\scriptsize] {$-.5$};
				\draw (.25,0.3pt) -- (.25,-1pt)
				node[anchor=north,font=\scriptsize] {$-.25$};
				\draw (.5,0.3pt) -- (.5,-1pt)
				node[anchor=north,font=\scriptsize] {$0$};
				\draw (.75,0.3pt) -- (.75,-1pt)
				node[anchor=north,font=\scriptsize] {$.25$};
				\draw (1,0.3pt) -- (1,-1pt)
				node[anchor=north,font=\scriptsize] {$.5$};
				\draw (0.3pt,0) -- (-0.8pt,0);
				\draw (0.3pt,{1/5}) -- (-0.8pt,{1/5});
				\draw (0.3pt,{2/5}) -- (-0.8pt,{2/5});
				\draw (0.3pt,{3/5}) -- (-0.8pt,{3/5});
				\draw (0.3pt,{4/5}) -- (-0.8pt,{4/5});
				\draw (0.3pt,1) -- (-0.8pt,1);
				\node[left,font=\scriptsize] at (0,0) {$0$};
				\node[left,font=\scriptsize] at (0,{1/5}) {$.2$};
				\node[left,font=\scriptsize] at (0,{2/5}) {$.4$};
				\node[left,font=\scriptsize] at (0,{3/5}) {$.6$};
				\node[left,font=\scriptsize] at (0,{4/5}) {$.8$};
				\node[left,font=\scriptsize] at (0,1) {$1$};		
				\draw[red!0!blue] plot[smooth,mark=square,mark indices={3,7,14,16},mark size=.8pt] file {power2bc1_rho2.txt};
				\draw[red!50!blue] plot[smooth,mark=triangle,mark indices={2,4,6,8,14,16,18,20},mark size=1pt] file {power2bc2_rho2.txt};	
				\draw[red!100!blue] plot[smooth,mark=asterisk,mark indices={6,8,15,19},mark size=1pt] file {power2dh_rho2.txt};
				\node[right] at (1.1,0) {$\gamma$};
			\end{tikzpicture}
			\begin{tikzpicture}[scale=3.5]
				\tikzstyle{vertex}=[font=\small,circle,draw,fill=yellow!20]
				\tikzstyle{edge} = [font=\scriptsize,draw,thick,-]
				\draw[black, thick] (0,0) -- (0,1);
				\draw[black, thick, ->] (0,0) -- (1.1,0);
				\draw (0,0.3pt) -- (0,-1pt)
				node[anchor=north,font=\scriptsize] {$-.5$};
				\draw (.25,0.3pt) -- (.25,-1pt)
				node[anchor=north,font=\scriptsize] {$-.25$};
				\draw (.5,0.3pt) -- (.5,-1pt)
				node[anchor=north,font=\scriptsize] {$0$};
				\draw (.75,0.3pt) -- (.75,-1pt)
				node[anchor=north,font=\scriptsize] {$.25$};
				\draw (1,0.3pt) -- (1,-1pt)
				node[anchor=north,font=\scriptsize] {$.5$};
				\draw (0.3pt,0) -- (-0.8pt,0);
				\draw (0.3pt,{1/5}) -- (-0.8pt,{1/5});
				\draw (0.3pt,{2/5}) -- (-0.8pt,{2/5});
				\draw (0.3pt,{3/5}) -- (-0.8pt,{3/5});
				\draw (0.3pt,{4/5}) -- (-0.8pt,{4/5});
				\draw (0.3pt,1) -- (-0.8pt,1);
				\node[left,font=\scriptsize] at (0,0) {$0$};
				\node[left,font=\scriptsize] at (0,{1/5}) {$.2$};
				\node[left,font=\scriptsize] at (0,{2/5}) {$.4$};
				\node[left,font=\scriptsize] at (0,{3/5}) {$.6$};
				\node[left,font=\scriptsize] at (0,{4/5}) {$.8$};
				\node[left,font=\scriptsize] at (0,1) {$1$};		
				\draw[red!0!blue] plot[smooth,mark=square,mark indices={3,7,9,14,16},mark size=.8pt] file {power2bc1_rho3.txt};
				\draw[red!50!blue] plot[smooth,mark=triangle,mark indices={4,6,8,14,16,18},mark size=1pt] file {power2bc2_rho3.txt};	
				\draw[red!100!blue] plot[smooth,mark=asterisk,mark indices={6,8,13,15,19},mark size=1pt] file {power2dh_rho3.txt};
				\node[right] at (1.1,0) {$\gamma$};
			\end{tikzpicture}
			\caption{Alternative rejection frequencies using matched pair samples of size $n=500$ and nominal level $\alpha=.05$. The left, center and right panels correspond to $\rho=.25,.5,.75$. Rejection frequencies are plotted for the modified WMW test with $\tau_n=.75$ (superimposed with squares) and with $\tau_n-\infty$ (superimposed with trangles), and the DH test with tuning parameter $-.139$ (superimposed with asterisks). The P-P curves, parametrized by $\gamma$, are displayed in top-left and bottom-left panels of \cref{fig:power}.}\label{fig:power_pairs}
		\end{figure}
		
		\section{Empirical illustration}
		\label{sec:empirical}
		
		We illustrate the modified WMW test of first-order stochastic dominance with an application to Canadian family income distributions. Our dataset is the same as the one used in \citet{BD03} and \citet{DH16}, so our results may be compared directly to those reported there. The data consist of before-tax and after-tax family incomes in the Canadian Family Expenditure Survey for the years 1978 and 1986. There are 8526 observations in the former year and 9470 in the latter. We test two null hypotheses: that the income distribution in 1986 first-order stochastically dominates the income distribution in 1978 (abbreviated as $1986\gtrsim1978$), and that the income distribution in 1978 first-order stochastically dominates the income distribution in 1986 ($1978\gtrsim1986$), reporting results for before-tax and after-tax income separately.
		
		In \cref{fig:empirical} we display the P-P plots for before-tax and after-tax incomes. The two plots are similar. We see that they lie neither entirely above nor entirely below the 45-degree line, indicating the possibility that both null hypotheses may be rejected. In particular, the P-P plots show that the estimated 30th percentile of the income distribution, either before-tax or after-tax, declined slightly between 1978 and 1986. The population P-P curves must be everywhere no greater than the 45-degree line if $1986\gtrsim1978$, or everywhere no less than the 45-degree line if $1978\gtrsim1986$.
		
		\begin{figure}[]
			\centering
				\begin{tikzpicture}[scale=3.5]
					\tikzstyle{vertex}=[font=\small,circle,draw,fill=yellow!20]
					\tikzstyle{edge} = [font=\scriptsize,draw,thick,-]
					\draw[black, thick] (0,0) -- (0,1);
					\draw[black, thick] (0,0) -- (1,0);
					\draw (0,0.3pt) -- (0,-1pt)
					node[anchor=north,font=\scriptsize] {$0$};
					\draw (.2,0.3pt) -- (.2,-1pt)
					node[anchor=north,font=\scriptsize] {$.2$};
					\draw (.4,0.3pt) -- (.4,-1pt)
					node[anchor=north,font=\scriptsize] {$.4$};
					\draw (.6,0.3pt) -- (.6,-1pt)
					node[anchor=north,font=\scriptsize] {$.6$};
					\draw (.8,0.3pt) -- (.8,-1pt)
					node[anchor=north,font=\scriptsize] {$.8$};
					\draw (1,0.3pt) -- (1,-1pt)
					node[anchor=north,font=\scriptsize] {$1$};
					\draw (0.3pt,0) -- (-0.8pt,0);
					\draw (0.3pt,{1/5}) -- (-0.8pt,{1/5});
					\draw (0.3pt,{2/5}) -- (-0.8pt,{2/5});
					\draw (0.3pt,{3/5}) -- (-0.8pt,{3/5});
					\draw (0.3pt,{4/5}) -- (-0.8pt,{4/5});
					\draw (0.3pt,1) -- (-0.8pt,1);
					\node[left,font=\scriptsize] at (0,0) {$0$};
					\node[left,font=\scriptsize] at (0,{1/5}) {$.2$};
					\node[left,font=\scriptsize] at (0,{2/5}) {$.4$};
					\node[left,font=\scriptsize] at (0,{3/5}) {$.6$};
					\node[left,font=\scriptsize] at (0,{4/5}) {$.8$};
					\node[left,font=\scriptsize] at (0,1) {$1$};			
					\draw[red!0!blue,thick,dotted,domain=0:1] plot(\x, \x);
					\draw[red!100!blue] plot file {Rbtplot.txt};
				\end{tikzpicture}	
				\begin{tikzpicture}[scale=3.5]
					\tikzstyle{vertex}=[font=\small,circle,draw,fill=yellow!20]
					\tikzstyle{edge} = [font=\scriptsize,draw,thick,-]
					\draw[black, thick] (0,0) -- (0,1);
					\draw[black, thick] (0,0) -- (1,0);
					\draw (0,0.3pt) -- (0,-1pt)
					node[anchor=north,font=\scriptsize] {$0$};
					\draw (.2,0.3pt) -- (.2,-1pt)
					node[anchor=north,font=\scriptsize] {$.2$};
					\draw (.4,0.3pt) -- (.4,-1pt)
					node[anchor=north,font=\scriptsize] {$.4$};
					\draw (.6,0.3pt) -- (.6,-1pt)
					node[anchor=north,font=\scriptsize] {$.6$};
					\draw (.8,0.3pt) -- (.8,-1pt)
					node[anchor=north,font=\scriptsize] {$.8$};
					\draw (1,0.3pt) -- (1,-1pt)
					node[anchor=north,font=\scriptsize] {$1$};
					\draw (0.3pt,0) -- (-0.8pt,0);
					\draw (0.3pt,{1/5}) -- (-0.8pt,{1/5});
					\draw (0.3pt,{2/5}) -- (-0.8pt,{2/5});
					\draw (0.3pt,{3/5}) -- (-0.8pt,{3/5});
					\draw (0.3pt,{4/5}) -- (-0.8pt,{4/5});
					\draw (0.3pt,1) -- (-0.8pt,1);
					\node[left,font=\scriptsize] at (0,0) {$0$};
					\node[left,font=\scriptsize] at (0,{1/5}) {$.2$};
					\node[left,font=\scriptsize] at (0,{2/5}) {$.4$};
					\node[left,font=\scriptsize] at (0,{3/5}) {$.6$};
					\node[left,font=\scriptsize] at (0,{4/5}) {$.8$};
					\node[left,font=\scriptsize] at (0,1) {$1$};		
					\draw[red!0!blue,dotted,thick,domain=0:1] plot(\x, \x);
					\draw[red!100!blue] plot file {Ratplot.txt};
				\end{tikzpicture}	
			\caption{P-P plots for Canadian family income before (left) and after (right) tax. The horizontal axes correspond to the year 1978, and the vertical axes to 1986. The 45-degree line is displayed as a dotted line in each panel.}\label{fig:empirical}
		\end{figure}
		
		In \cref{tab:empirical} we report the p-values of tests of first-order stochastic dominance. The first two rows of p-values, for the Barrett-Donald (BD) and DH tests, are taken directly from Tables 6 and 7 in \citet{DH16}, and confirmed to be correct (up to random variation intrinsic to the bootstrap) in independent calculation. The DH test uses a tuning parameter value of $-.15\approx-.1\sqrt{\log\log(n_1+n_2)}$. To obtain a tuning parameter value for the modified WMW test, we observe that setting $\tau_n=.5$ was effective in controlling the false rejection rate at the least favorable case in the simulations reported in \cref{lfctab} for the largest sample size of 1000 observations per sample. Since $.5\approx.35\sqrt{\log\log(n_1+n_2)}$ when $n_1=n_2=1000$, and $.53\approx.35\sqrt{\log\log(n_1+n_2)}$ when $n_1=8526$ and $n_2=9470$, we set $\tau_n=.53$ in this illustration. Similar to what is done in \citet{DH16}, to reduce computational burden we computed one-sided WMW statistics using a discrete approximation based on 2000 grid points spread evenly over the domain of the P-P plots. We calculated p-values using $10^5$ randomly generated bootstrap samples.		
		
		\begin{table}
			\centering
			\begin{threeparttable}
				\caption{p-values of stochastic dominance tests for Canadian family income.}\label{tab:empirical}
				{\small
				\begin{tabular}{lcccc}
					\toprule
					&\multicolumn{2}{c}{Before tax}&\multicolumn{2}{c}{After tax}\\
					\cmidrule(lr){2-3}\cmidrule(lr){4-5}
					Test & $1986\gtrsim1978$ & $1978\gtrsim1986$ & $1986\gtrsim1978$ & $1978\gtrsim1986$\\
					\midrule
					Barrett-Donald&0.0133&0.0000&0.0049&0.0009\\
					Donald-Hsu&0.0088&0.0000&0.0044&0.0003\\
					Wilcoxon-Mann-Whitney (standard)&0.1894&0.0022&0.0639&0.0539\\
					Wilcoxon-Mann-Whitney (modified)&0.0394&0.0000&0.0130&0.0003\\
					Linton-Song-Whang (original data)&0.1010&0.0000&0.0310&0.0000\\
					Linton-Song-Whang (squared data)&0.2700&0.0000&0.0870&0.0010\\
					Linton-Song-Whang (square-root data)&0.0750&0.0000&0.0220&0.0020\\
					\bottomrule
				\end{tabular}
				}
			\end{threeparttable}
		\end{table}
		
		We see in \cref{tab:empirical} that our modification to bootstrap critical values leads to a meaningful reduction in the p-values obtained using the one-sided WMW statistic. With the before-tax and after-tax incomes, the null hypothesis $1986\gtrsim1978$ is not rejected at the 5\% nominal level using the standard bootstrap critical value, but is rejected using the modified bootstrap critical value. The same is true for the null hypothesis $1978\gtrsim1986$ using the after-tax incomes, and in this case the modified bootstrap critical value also leads to rejection at the 1\% (or even 0.1\%) nominal level. Both the standard and modified bootstrap critical values lead to rejection of the null hypothesis $1978\gtrsim1986$ at the 1\% nominal level using the before-tax incomes.
		
		The p-values obtained using the one-sided WMW statistic with the standard bootstrap critical value are larger than those obtained using the one-sided Kolmogorov-Smirnov statistic with the standard bootstrap critical value (i.e., the BD test), and the p-values obtained using the one-sided WMW statistic with the modified bootstrap critical value are larger than those obtained using the one-sided Kolmogorov-Smirnov statistic with the modified bootstrap critical value (i.e., the DH test). This therefore appears to be a case where the particular way in which first-order stochastic dominance is violated is more easily detected with the one-sided Kolmogorov-Smirnov statistic than with the one-sided WMW statistic.
		
		The final three rows of \cref{tab:empirical} report p-values for the LSW test with $\kappa_n=.07\approx3(2T_n)^{-1/2}\log\log(2T_n)$. The first of the three rows shows the p-values obtained using the original income data, while the second and third show those obtained when the square or square-root function is applied to incomes. As discussed in \cref{sec:LSW}, the LSW test of first-order stochastic dominance is not invariant to strictly increasing transformations of the data, so applying the square or square-root function to incomes can affect the p-value obtained. Indeed, the final three rows of \cref{tab:empirical} differ substantially, and the outcome of the LSW test of $1986\gtrsim1978$ switches from rejection to non-rejection at the $5\%$ nominal level after applying the square function to after-tax incomes. Transforming incomes using the square or square-root function does not affect the p-values reported in the first four rows of \cref{tab:empirical} because the corresponding tests are invariant to strictly increasing transformations of the data.
		
		\section{Final remarks}
		\label{sec:final}
		
		We have shown in this article that the bootstrap can be used to implement tests of first-order stochastic dominance based on the one-sided WMW statistic in settings involving either independent sampling or matched pair sampling. Moreover, we have shown that modifying the bootstrap so that it incorporates an estimate of the contact set can lead to a large improvement in power while maintaining control of false rejection rates. Our procedure can be applied alongside existing tests using other statistics, such as the LSW test and the DH test, in routine applications of stochastic dominance testing. As tests based on different statistics may differ in their propensity to detect different violations of stochastic dominance, it can be informative to compare the outcome of multiple tests.
		
		Versions of the LSW and DH tests suitable for testing second-order stochastic dominance are provided in \citet{LSW10} and \cite{DH16}. It would be useful to extend the methods proposed in this article so as to likewise obtain a test of second-order stochastic dominance. A complicating factor is the central role played by the P-P process and bootstrap P-P process in our asymptotic arguments, and our reliance on the representation of the one-sided WMW statistic as a functional of the P-P plot. The P-P plot is not a suitable tool for assessing second-order stochastic dominance because such dominance is not invariant under strictly increasing transformations and so cannot be expressed as a property of the P-P curve. It has recently been shown in \citet{LL25} that second-order stochastic dominance can instead be expressed as a property of the Lorenz P-P curve. The Lorenz P-P curve is a nondecreasing function from $[0,1]$ into $[0,1]$ obtained by composing the inverse of one unscaled Lorenz curve with another unscaled Lorenz curve, and is everywhere no greater than the 45-degree line precisely when second-order stochastic dominance is satisfied. Bootstrap tests of second-order stochastic dominance proposed in \citet{LL25} are similar in spirit to those proposed in \citet{BD03} and do not make use of a contact set estimator. It seems likely that the power of the Lando-Legramanti tests could be improved by incorporating contact set estimation, similar to what has been done here and in \citet{LSW10}. We leave this as a topic for future research.

		\section{Proofs}
		\label{sec:proofs}
		
		\cref{prop:F} is proved in \cref{sec:proofWMW}. \cref{prop:tauinfinity,prop:modtest} are proved in \cref{sec:proofbs}. \cref{prop:L1odc,prop:bswc} are proved in \citet{BK26}. \cref{sec:FS,sec:proofnondeg} establish auxiliary lemmas respectively concerning the consistency of the modified bootstrap and the regularity of asymptotic distributions.		
		
		\subsection{Convergence in distribution of the one-sided Wilcoxon-Mann-Whitney statistic}
		\label{sec:proofWMW}
		
		With \cref{prop:L1odc} in hand, the proof of \cref{prop:F} is a straightforward application of the delta-method for Hadamard \emph{directionally} differentiable maps. The latter property is weaker than Hadamard differentiability because the derivative is not required to be linear. See Definition 2.1 and Theorem 2.1 in \citet{FS19} for the definition of Hadamard directional differentiability and a corresponding statement of the delta-method. The idea originates in \citet{S90,S91} and \citet{D93}. See also \citet{BM15} and \citet{K16}.
		
		\begin{lemma}\label{lemma:HDD}
			The map $\mathcal{H}:L^1[0,1]\to \mathbb{R}$ defined in \eqref{eq:H} is Hadamard
			directionally differentiable at each $\theta\in L^1[0,1]$, with directional derivative $\mathcal H'_\theta:L^1[0,1]\to\mathbb R$ given by
			\begin{equation*}
				\mathcal H'_\theta(h)=\int_0^1\mathbbm 1(\theta(u)>u)h(u)\,\mathrm{d}u+\int_0^1\mathbbm 1(\theta(u)=u)\max\{h(u),0\}\,\mathrm{d}u.
			\end{equation*}
		\end{lemma}
		\begin{proof}
			Let $\{t_n\}$ be a sequence of positive real numbers such that $t_n\to0$. Let $\{h_n\}$ be a sequence in $L^1[0,1]$ converging to $h\in L^1[0,1]$. Let $\theta\in L^1[0,1]$. Our task is to show that
			\begin{equation*}
				t_n^{-1}[\mathcal H(\theta+t_nh_n)-\mathcal H(\theta)]\to\int_0^1\mathbbm 1(\theta(u)>u)h(u)\,\mathrm{d}u+\int_0^1\mathbbm 1(\theta(u)=u)\max\{h(u),0\}\,\mathrm{d}u.
			\end{equation*}
			From the definition of $\mathcal H$ we have
			\begin{align*}
				t_n^{-1}[\mathcal H(\theta+t_nh_n)-\mathcal H(\theta)]&=\int_0^1\mathbbm 1(\theta(u)>0)\,\Big(\max\{h(u)+t_n^{-1}(\theta(u)-u),0\}-\max\{t_n^{-1}(\theta(u)-u),0\}\Big)\,\mathrm{d}u\\
				&\quad+\int_0^1\mathbbm 1(\theta(u)=0)\max\{h(u),0\}\,\mathrm{d}u\\
				&\quad+\int_0^1\mathbbm 1(\theta(u)<0)\,\Big(\max\{h(u)+t_n^{-1}(\theta(u)-u),0\}-\max\{t_n^{-1}(\theta(u)-u),0\}\Big)\,\mathrm{d}u\\
				&\quad+\int_0^1\Big(\max\{h_n(u)+t_n^{-1}(\theta(u)-u),0\}-\max\{h(u)+t_n^{-1}(\theta(u)-u),0\}\Big)\,\mathrm{d}u.
			\end{align*}
			For every two real numbers $x$ and $y$ we have $\lvert\max\{x,0\}-\max\{y,0\}\rvert\leq\lvert x-y\rvert$. Therefore the first and third integrands are bounded in magnitude by $\lvert h(u)\rvert$ and the fourth integrand is bounded in magnitude by $\lvert h_n(u)-h(u)\rvert$. Consequently the first integral converges to $\smallint_0^1\mathbbm 1(\theta(u)>0)h(u)\,\mathrm{d}u$ by the dominated convergence theorem, the third integral converges to zero by the dominated convergence theorem, and the fourth integral converges to zero because $\lVert h_n-h\rVert_1\to0$.\quad\raisebox{-1pt}{$\Box$}
		\end{proof}
		
		Note that the directional derivative $\mathcal H'_\theta$ is not linear unless $\theta(u)\neq u$ for a.e.\ $u\in[0,1]$. Thus $\mathcal H$ is not Hadamard differentiable everywhere on $L^1[0,1]$. It is Hadamard directionally differentiable everywhere on $L^1[0,1]$.
		
		\begin{proof}[Proof of \cref{prop:F}]
			The first assertion follows from \cref{prop:L1odc} and \cref{lemma:HDD} by applying the delta-method with the map $\mathcal H$. The second and third assertions follow from the first since $\mathcal H(R)=0$ under $\mathrm{H}_0$, $\mathcal H(R)>0$ under $\mathrm{H}_1$, and $\hat{S}=T^{1/2}_n\mathcal H(\hat{R})+O(n_2^{-1/2})$ by \eqref{eq:approx}.\quad\raisebox{-1pt}{$\Box$}
		\end{proof}
		
		\subsection{Consistency of the modified bootstrap}
		\label{sec:FS}
		
		The following lemma establishing good behavior of the modified bootstrap test statistic $\tilde{S}^\ast$ will be used in the proof of \cref{prop:modtest}.
		
		\begin{lemma}\label{lem:FS}
			If \cref{ass:distribution,ass:data,ass:nondegenerate,ass:tau} are satisfied and if $\mathrm{H}_0$ is true then $\tilde{S}^\ast\convd\mathcal H'_R(\mathcal R)$ conditional on the data in probability.
		\end{lemma}
		
		We will prove \cref{lem:FS} by applying Theorem 3.2 in \citet{FS19}, which provides general conditions under which bootstrap procedures based on an estimated contact set are well-behaved. The proof will also make use of the following two lemmas.
		
		\begin{lemma}\label{lem:pointwiseOp}
			If \cref{ass:data} is satisfied then $T_n^{1/2}(\hat R(u)-R(u))=O_{\mathrm{P}}(1)$ for every $u\in(0,1)$ at which $R$ is differentiable.
		\end{lemma}
		\begin{proof}
			For each $u\in(0,1)$ we have
			\begin{align*}
				T^{1/2}_n(\hat R(u)-R(u))=\sqrt{\frac{n_2}{n_1+n_2}}n_1^{1/2}(\hat F_1(\hat Q_2(u))-F_1(\hat Q_2(u)))+\sqrt{\frac{n_1}{n_1+n_2}}n_2^{1/2}(F_1(\hat Q_2(u))-R(u)).
			\end{align*}
			The term $n_1^{1/2}(\hat F_1(\hat Q_2(u))-F_1(\hat Q_2(u)))$ is $O_{\mathrm{P}}(1)$ by Donsker's theorem. Therefore, since $n_2/(n_1+n_2)\to\lambda$, it suffices to show that $\displaystyle{n_2^{1/2}(F_1(\hat{Q}_2(u))-R(u))=O_{\mathrm P}(1)}$ if $u$ is a differentiability point of $R$.
			
			Let $\{Y_i\}_{i=1}^{n_2}$ be iid random variables uniformly distributed on $(0,1)$, and let $\hat{U}$ be their empirical quantile function. The iid random variables $\{Q_2(Y_i)\}_{i=1}^{n_2}$ have quantile function $Q_2$ and empirical quantile function $Q_2\circ\hat{U}$. Thus $\hat{Q}_2(u)\eqd Q_2(\hat{U}(u))$. Consequently $F_1(\hat{Q}_2(u))\eqd R(\hat{U}(u))$. As is well-known, $\displaystyle{n_2^{1/2}(\hat{U}(u)-u)\convd N(0,u(1-u))}$; see, for instance, Example 21.6 in \citet{V98}. If $R$ is differentiable at $u$ then an application of the delta-method shows that $n_2^{1/2}(R(\hat{U}(u))-R(u))\convd N(0,R'(u)^2u(1-u))$. Thus $n_2^{1/2}(F_1(\hat{Q}_2(u))-R(u))\eqd n_2^{1/2}(R(\hat{U}(u))-R(u))=O_{\mathrm{P}}(1)$.\quad\raisebox{-1pt}{$\Box$}
		\end{proof}
		
		\begin{lemma}\label{lem:copula}
			Suppose that \cref{ass:data}\textup{(ii)} is satisfied. Let $(u,v)\in\ran(F_1)\times\ran(F_2)$ be such that $\mathrm{P}(F_1(X_i^1)=u)=0$ and $\mathrm{P}(F_2(X_i^2)=v)=0$. Let $\{(u_n,v_n)\}_{n=1}^\infty$ be a sequence in $[0,1]^2$ such that $(u_n,v_n)\to(u,v)$. Then $\hat C(u_n,v_n)\convp C(u,v)$.
		\end{lemma}
		\begin{proof}
			In this proof we let $\hat{U}_i=\hat{F}_1(X_i^1)$, $\hat{V}_i=\hat{F}_2(X_i^2)$, $U_i=F_1(X_i^1)$ and $V_i=F_2(X_i^2)$. Fix $\epsilon>0$ and use Markov's inequality to obtain
			\begin{equation*}
				\mathrm{P}\,\left(\frac{1}{n}\sum_{i=1}^n\left|\mathbbm 1(\hat{U}_i\leq u_n,\hat{V}_i\leq v_n)-\mathbbm 1(U_i\leq u,V_i\leq v)\right|>\epsilon\right)\leq\frac{1}{n\epsilon}\sum_{i=1}^n\mathrm{E}\left|\mathbbm 1(\hat{U}_i\leq u_n,\hat{V}_i\leq v_n)-\mathbbm 1(U_i\leq u,V_i\leq v)\right|.
			\end{equation*}
			The expected value does not vary with $i$ because the sample $\{(X_i^1,X_i^2)\}_{i=1}^n$ is iid, thus exchangeable. Therefore
			\begin{equation}
				\mathrm{P}\,\left(\frac{1}{n}\sum_{i=1}^n\left|\mathbbm 1(\hat{U}_i\leq u_n,\hat{V}_i\leq v_n)-\mathbbm 1(U_i\leq u,V_i\leq v)\right|>\epsilon\right)\leq\frac{1}{\epsilon}\,\mathrm{E}\left|\mathbbm 1(\hat{U}_1\leq u_n,\hat{V}_1\leq v_n)-\mathbbm 1(U_1\leq u,V_1\leq v)\right|.\label{eq:copulaMarkov}
			\end{equation}
			The Glivenko-Cantelli theorem shows that $(\hat{U}_1,\hat{V}_1)\convp(U_1,V_1)$. Therefore $(\hat{U}_1+u-u_n,\hat{V}_1+v-v_n)\convp(U_1,V_1)$. The map $f_{u,v}:[0,1]^2\to\{0,1\}$ defined by $f_{u,v}(a,b)=\mathbbm 1(a\leq u,b\leq v)$ is continuous on a subset of $[0,1]^2$ inhabited by $(U_1,V_1)$ with probability one under our assumption that $\mathrm{P}(U_1=u)=0$ and $\mathrm{P}(V_1=v)=0$. Therefore an application of the continuous mapping theorem shows that $Z_n\coloneqq |f_{u,v}(\hat{U}_1+u-u_n,\hat{V}_1+v-v_n)-f_{u,v}(U_1,V_1)|\convp0$. The random variables $Z_n$ are uniformly integrable because they are nonnegative and bounded by one. Thus $\mathrm{E}(Z_n)\to 0$. Since $Z_n=|\mathbbm 1(\hat{U}_1\leq u_n,\hat{V}_1\leq v_n)-\mathbbm 1(U_1\leq u,V_1\leq v)|$, we deduce from \eqref{eq:copulaMarkov} that
			\begin{equation*}
				\frac{1}{n}\sum_{i=1}^n\left|\mathbbm 1(\hat{U}_i\leq u_n,\hat{V}_i\leq v_n)-\mathbbm 1(U_i\leq u,V_i\leq v)\right|\convp0.
			\end{equation*}
			Consequently $\hat{C}(u_n,v_n)=n^{-1}\sum_{i=1}^n\mathbbm 1(U_i\leq u,V_i\leq v)+o_{\mathrm{P}}(1)$. We have $n^{-1}\sum_{i=1}^n\mathbbm 1(U_i\leq u,V_i\leq v)\convp\mathrm{P}(U_i\leq u,V_i\leq v)$ by the law of large numbers. It remains only to show that $C(u,v)=\mathrm{P}(U_i\leq u,V_i\leq v)$. Let $x_1$ and $x_2$ be real numbers such that $F_1(x_1)=u$ and $F_2(x_2)=v$. Then $C(u,v)=C(F_1(x_1),F_2(x_2))=\mathrm{P}(X_i^1\leq x_1,X_i^2\leq x_2)$. Basic properties of quantile functions show that $Q_1(U_i)=X_i^1$ a.s., that $Q_2(V_i)=X_i^2$ a.s., that $Q_1(U_i)\leq x_1$ if and only if $U_i\leq F_1(x_1)$, and that $Q_2(V_i)\leq x_2$ if and only if $V_i\leq F_2(x_2)$. Therefore $\mathrm{P}(X_i^1\leq x_1,X_i^2\leq x_2)=\mathrm{P}(U_i\leq u,V_i\leq v)$.\quad\raisebox{-1pt}{$\Box$}
		\end{proof}
		
		\begin{proof}[Proof of \cref{lem:FS}]
			In view of \eqref{eq:implicit} it suffices for us to verify Assumptions 1--4 of Theorem 3.2 in \citet{FS19} with $\hat{\phi}'_n=\hat{\mathcal H}'$, $\phi=\mathcal H$, $\hat{\theta}_n^\ast=\hat{R}^\ast$, $\hat{\theta}_n=\hat{R}$, $\theta_0=R$, $\displaystyle{r_n=T_n^{1/2}}$ and $\mathbb G_0=\mathcal R$. Assumption 1 is satisfied by \cref{lemma:HDD}. Assumption 2 is satisfied by \cref{prop:L1odc}, noting that every probability measure on a separable Banach space is tight. Assumption 3 is satisfied by \cref{prop:bswc}, noting that $\hat R$ and $\hat R^\ast$ are Borel measurable maps from the underlying probability space into $L^1[0,1]$.
			
			For Assumption 4 of Theorem 3.2 in \citet{FS19} to be satisfied, it suffices to show (see Remark 3.4 therein) that $\hat{\mathcal H}'(h)\convp\mathcal H'_R(h)$ for every $h\in L^1[0,1]$. Observe that $\lvert\hat{\mathcal H}'(h)-\mathcal H'_R(h)\rvert\leq\smallint_{\hat{B}_0\triangle B_0}\max\{h(u),0\}\,\mathrm{d}u$, where $\triangle$ is the symmetric difference of sets. By applying Markov's inequality and Fubini's theorem we find that, for every $\epsilon>0$,
			\begin{equation*}
				\mathrm{P}\left(\lvert\hat{\mathcal H}'(h)-\mathcal H'_R(h)\rvert>\epsilon\right)\leq\frac{1}{\epsilon}\,\mathrm{E}\,\lvert\hat{\mathcal H}'(h)-\mathcal H'_R(h)\rvert
				\leq\frac{1}{\epsilon}\int_0^1\mathrm{P}\big(u\in\hat{B}_0\triangle B_0\big)\max\{h(u),0\}\,\mathrm{d}u.
			\end{equation*}
			Assumption 4 therefore follows from the dominated convergence theorem if
			\begin{equation}\label{eq:symdif}
				\mathrm{P}\,(u\in\hat{B}_0\triangle B_0)\to0\quad\text{ for a.e.\ }u\in(0,1).
			\end{equation}
			In particular, it suffices to show that the convergence holds for every $u\in(0,1)$ such that
			\begin{enumerate}[label=\upshape(\roman*)]
				\item $R$ is differentiable at $u$;
				\item $\mathrm{P}(F_1(X_i^1)=u)=0$ and $\mathrm{P}(F_2(X_i^2)=u)=0$;
				\item if $u\in\ran(F_1)\cap\ran(F_2)$ then $C(u,u)<u$; and
				\item if $u\in B_0$ then $u\in\ran(F_2)$.
			\end{enumerate}
			Note that (i) is satisfied for a.e.\ $u\in(0,1)$ under \cref{ass:distribution}; that (ii) is satisfied for a.e.\ $u\in(0,1)$ because every probability measure on $[0,1]$ has at most countably many mass points; that (iii) is satisfied for a.e.\ $u\in(0,1)$ under \cref{ass:nondegenerate}; and that (iv) is satisfied for a.e.\ $u\in(0,1)$ because the complement to $\ran(F_2)$ is a countable union of intervals over which $R$ is constant, so that at most one point in each interval belongs to $B_0$. Fix a point $u\in(0,1)$ satisfying conditions (i)--(iv) in what follows.
			
			Suppose that $u\in B_0$. In this case we have $u\in\ran(F_1)$ because $u=R(u)\in\ran(R)\subseteq\ran(F_1)$, and we have
			\begin{equation}\label{eq:uinB0}
				\mathrm{P}\,(u\in\hat{B}_0\triangle B_0)=\mathrm{P}(u\notin\hat{B}_0)=\mathrm{P}\left(T_n^{1/2}(\hat R(u)-R(u))\leq-\tau_n\hat V^{1/2}_{\lceil n_2u\rceil}\right).
			\end{equation}
			\cref{lem:pointwiseOp} establishes that $T_n^{1/2}(\hat R(u)-R(u))=O_{\mathrm{P}}(1)$ under condition (i), and \cref{ass:tau} requires that $\tau_n\to\infty$. Therefore if $\hat{V}_{\lceil n_2u\rceil}$ converges in probability to a positive constant then the final probability in \eqref{eq:uinB0} converges to zero. For the independent sampling framework we trivially have $\hat{V}_{\lceil n_2u\rceil}\to u(1-u)>0$. For the matched pairs sampling framework we have
			\begin{equation*}
				\hat{V}_{\lceil n_2u\rceil}=\frac{\lceil nu\rceil}{n}-\hat{C}\left(\frac{\lceil nu\rceil}{n},\frac{\lceil nu\rceil}{n}\right)\convp u-C(u,u)
			\end{equation*}
			by \cref{lem:copula} under conditions (ii) and (iv). Conditions (iii) and (iv) together ensure that $u-C(u,u)>0$. Therefore $\hat{V}_{\lceil n_2u\rceil}$ also converges in probability to a positive constant under matched pairs sampling.
			
			Suppose instead that $u\notin B_0$. In this case we have
			\begin{equation}\label{eq:unotinB}
				\mathrm{P}\,(u\in\hat{B}_0\triangle B_0)=\mathrm{P}(u\in\hat{B}_0)=\mathrm{P}\left(\hat R(u)-R(u)+T_n^{-1/2}\tau_n\hat V^{1/2}_{\lceil n_2u\rceil}> u-R(u)\right).
			\end{equation}
			We have $\hat R(u)-R(u)\convp0$ by \cref{lem:pointwiseOp} under condition (i), and we have $T_n^{-1/2}\tau_n\hat{V}_{\lceil n_2u\rceil}^{1/2}\to0$ because $\displaystyle{T_n^{-1/2}\tau_n\to0}$ under \cref{ass:tau} and because $\lvert\hat{V}_{\lceil n_2u\rceil}\rvert\leq1$. We also have $u-R(u)>0$ because of the assumed truth of $\mathrm{H}_0$ and because $u\notin B_0$. Therefore the final probability in \eqref{eq:unotinB} converges to zero. This shows that \eqref{eq:symdif} is satisfied and completes our verification of Assumptions 1-4 in \citet{FS19}.\quad\raisebox{-1pt}{$\Box$}
		\end{proof}
		
		\subsection{Regularity of asymptotic distributions}
		\label{sec:proofnondeg}
		
		A technical obstacle to proving \cref{prop:tauinfinity,prop:modtest} is the need to demonstrate that the quantile function for the asymptotic distribution of the bootstrap test statistic is continuous and strictly increasing at $1-\alpha$, where $\alpha\in(0,1/2)$. \cref{lem:nondegenerate} supplies this ingredient. To prove \cref{lem:nondegenerate} we use a theorem on convex functionals of Gaussian processes in \cite{DLS98} to show that the only possible mass point for the asymptotic distribution is zero, with any remaining mass spread smoothly over the entire nonnegative halfline. A separate argument, given in the proof of \cref{lem:nondegenerate0}, shows that the mass at zero can be no greater than $1/2$. Closely related econometric applications of the relevant theorem in \cite{DLS98} appear in, for instance, \citet{A02}, \citet{CF05}, \citet{ACF06}, \citet{CH06} and \citet{CHT07}.

		\begin{lemma}\label{lem:nondegenerate}
			Suppose that \cref{ass:distribution,ass:nondegenerate} are satisfied. Let $I:[0,1]\to[0,1]$ be the identity map. Let $Q_I$ and $Q_R$ be the quantile functions for $\mathcal H'_{I}(\mathcal R)$ and $\mathcal H'_{R}(\mathcal R)$. Then $Q_I$ and $Q_R$ are continuous on $(0,1)$ and $Q_I$ is strictly increasing on $(1/2,1)$. If $\mathrm{H}_0$ is true and $B_0$ has positive measure then $Q_R$ is also strictly increasing on $(1/2,1)$.
		\end{lemma}
		
		The next lemma is used to prove \cref{lem:nondegenerate}.
		
		\begin{lemma}\label{lem:nondegenerate0}
			Suppose that $R$ is absolutely continuous and that $C(u,u)<u$ for a.e.\ $u\in\ran(R)$ such that $r(u)>0$. Then
			\begin{equation*}
				\mathrm{P}\left(\int_{A}\max\{\mathcal R(u),0\}\mathrm{d}u=0\right)\leq\frac{1}{2}
			\end{equation*}
			for every positive measure set $A\subseteq(0,1)$ such that $0<R(u)<1$ for all $u\in A$.
		\end{lemma}
		\begin{proof}
			We begin by showing that $\mathrm{Var}(\mathcal R(u))>0$ for a.e.\ $u\in A$. Recall the elementary inequality
			\begin{equation*}
				v\wedge w-vw\leq\sqrt{vw(1-v)(1-w)}\quad\text{for all }v,w\in(0,1),\text{ strictly if }v\neq w.
			\end{equation*}
			Also recall that $\mathcal R(u)=\lambda^{1/2}\mathcal B_1(R(u))-(1-\lambda)^{1/2}r(u)\mathcal B_2(u)$. Using \eqref{copulacov} we compute $\mathrm{Var}(\mathcal B_1(R(u)))=R(u)(1-R(u))$ and $\mathrm{Var}(\mathcal B_2(u))=u(1-u)$. Set $v=R(u)$ and $w=u$ in the elementary inequality to obtain
			\begin{equation*}
				R(u)\wedge u-R(u)u\leq\sqrt{\mathrm{Var}(\mathcal B_1(R(u)))\mathrm{Var}(\mathcal B_2(u))}\quad\text{for all }u\in A,\text{ strictly if }R(u)\neq u.
			\end{equation*}
			From the Fr\'{e}chet-Hoeffding upper bound $C(R(u),u)\leq R(u)\wedge u$ and from the assumption on $C$ we obtain
			\begin{equation*}
				C(R(u),u)-R(u)u\leq R(u)\wedge u-R(u)u\quad\text{for all }u\in A,\text{ strictly for a.e.\ }u\in A\text{ such that }R(u)=u\text{ and }r(u)>0.
			\end{equation*}
			Using \eqref{copulacov} we compute $\mathrm{Cov}(\mathcal B_1(R(u)),\mathcal B_2(u))=C(R(u),u)-R(u)u$. Thus we have shown that
			\begin{equation*}
				\mathrm{Cov}(\mathcal B_1(R(u)),\mathcal B_2(u))<\sqrt{\mathrm{Var}(\mathcal B_1(R(u)))\mathrm{Var}(\mathcal B_2(u))}\quad\text{for a.e.\ }u\in A\text{ such that }r(u)>0.
			\end{equation*}
			If $\mathrm{Var}(\mathcal R(u))=0$ and $r(u)>0$ then $\mathcal B_1(R(u))$ and $\mathcal B_2(u)$ are linearly dependent with weights $\lambda^{1/2}>0$ and $-(1-\lambda)^{1/2}r(u)<0$, and so their covariance must attain the upper bound provided by the Cauchy-Schwarz inequality. However we have shown that, for a.e.\ $u\in A$ such that $r(u)>0$, this bound is not attained. Therefore $\mathrm{Var}(\mathcal R(u))>0$ for a.e.\ $u\in A$ such that $r(u)>0$. Moreover, for all $u\in A$ such that $r(u)=0$, we have $\mathrm{Var}(\mathcal R(u))=\lambda\mathrm{Var}(\mathcal B_1(R(u)))>0$. Therefore $\mathrm{Var}(\mathcal R(u))>0$ for a.e.\ $u\in A$.
			
			The events $\{\mathcal R(u)<0\text{ for a.e.\ }u\in A\}$ and $\{\mathcal R(u)>0\text{ for a.e.\ }u\in A\}$ are disjoint because $A$ has positive measure, and are equally probable because $\mathcal R$ and $-\mathcal R$ have the same distribution. Thus each event has probability no greater than $1/2$. Consequently
			\begin{equation*}
				\mathrm P\,\bigg(\int_A\max\{\mathcal R(u),0\}\,\mathrm{d}u=0\bigg)\leq\frac{1}{2}+\mathrm{P}\,\bigg(\int_A\mathbbm 1(\mathcal R(u)=0)\,\mathrm{d}u>0\bigg).
			\end{equation*}
			The final probability is zero if $\mathrm{E}\smallint_A\mathbbm 1(\mathcal R(u)=0)\,\mathrm{d}u=0$. By Fubini's theorem, this is the case if $\mathrm{P}(\mathcal R(u)=0)=0$ for a.e.\ $u\in A$. This condition is satisfied because $\mathcal R$ is Gaussian and because $\mathrm{Var}(\mathcal R(u))>0$ for a.e.\ $u\in A$.\quad\raisebox{-1pt}{$\Box$}
		\end{proof}
		
		\begin{proof}[Proof of \cref{lem:nondegenerate}]
			Let $F_I$ and $F_R$ be the cdfs for $\mathcal H'_I(\mathcal R)$ and $\mathcal H'_R(\mathcal R)$. Since $\mathcal R$ is a Gaussian random element of $L^1[0,1]$ and $\mathcal H'_I$ and $\mathcal H'_R$ are continuous and convex, Theorem 11.1 in \citet{DLS98} establishes that $F_I$ and $F_R$ can only be discontinuous at $x_I=\inf\{x:F_I(x)>0\}$ and $x_R=\inf\{x:F_R(x)>0\}$ respectively. It further establishes that $F_I$ is strictly increasing on $(x_I,\infty)$ if $F_I(x_I)<1$, and that $F_R$ is strictly increasing on $(x_R,\infty)$ if $F_R(x_R)<1$. We deduce from these facts that $Q_I$ and $Q_R$ are continuous on $(0,1)$, that $Q_I$ is strictly increasing on $(F_I(x_I),1)$ if $F_I(x_I)<1$, and that $Q_R$ is strictly increasing on $(F_R(x_R),1)$ if $F_R(x_R)<1$.
			
			Let $V$ be the support of $\mathcal R$ in $L^1[0,1]$; i.e., the set of all $h\in L^1[0,1]$ such that $\mathrm{P}(\lVert\mathcal R-h\rVert_1<\epsilon)>0$ for all $\epsilon>0$. Problem 11.3 in \citet{DLS98} shows that $x_I=\inf_{h\in V}\mathcal H'_I(h)$ and $x_R=\inf_{h\in V}\mathcal H'_R(h)$. We have $0\in V$ because every centered Gaussian measure on a separable Banach space contains zero in its support \citep{V75}, and we have $\mathcal H'_I(0)=\mathcal H'_R(0)=0$. Since $\mathcal H'_I\geq0$ it follows that $x_I=0$. If $\mathrm{H}_0$ is true then $\mathcal H'_R\geq0$, and so $x_R=0$.
			
			The proof is complete if we can show that $F_I(0)\leq1/2$ and show that if $B_0$ has positive measure then $F_R(0)\leq1/2$. By applying \cref{lem:nondegenerate0} with $A=\{u\in(0,1):0<R(u)<1\}$, which has positive measure because $R$ is continuous and not identically equal to zero or one, we find that $F_I(0)\leq1/2$. If $B_0$ has positive measure then by applying \cref{lem:nondegenerate0} with $A=B_0\cap(0,1)$ we find that $F_R(0)\leq1/2$. Note that the condition placed on $C$ in \cref{lem:nondegenerate0} is satisfied under \cref{ass:distribution,ass:nondegenerate} because $\ran(R)\subseteq\ran(F_1)$ and because $r(u)=0$ for a.e.\ $u\notin\ran(F_2)$.\quad\raisebox{-1pt}{$\Box$}
		\end{proof}

		\subsection{Rejection probabilities using the standard and modified bootstrap critical values}
		\label{sec:proofbs}
		
		We conclude by proving \cref{prop:tauinfinity,prop:modtest}. The proofs of both results make use of the following lemma, which may be compared to Lemma 10.11(i) in \citet{K08}.
		
		\begin{lemma}\label{lem:bsQconsistent}
			Let $(\Omega,\mathcal F,\mathrm{P})$ be a probability space, let $X$ and $\{X_n\}_{n=1}^\infty$ be real-valued random variables on $\Omega$, and let $\{\mathcal F_n\}_{n=1}^\infty$ be sub-$\sigma$-algebras of $\mathcal F$. Let $Q:(0,1)\to\mathbb R$ be the quantile function for $X$. Assume that $\mathrm{E}(h(X_n)\mid\mathcal F_n)\convp\mathrm{E}(h(X))$ for every Lipschitz continuous function $h:\mathbb R\to[0,1]$. Then, for every $u\in(0,1)$ at which $Q$ is continuous,
			\begin{equation*}
				\inf\left\{x\in\mathbb R:\mathrm{P}(X_n\leq x\mid\mathcal F_n)\geq u\right\}\convp Q(u).
			\end{equation*}
			
		\end{lemma}
		\begin{proof}
			Fix $\epsilon>0$ and a point $u\in(0,1)$ at which $Q$ is continuous. Let $\hat{c}=\inf\{x\in\mathbb R:\mathrm{P}(X_n\leq x\mid\mathcal F_n)\geq u\}$, let $c=Q(u)$, and let $F$ be the cdf for $X$. The continuity of $Q$ at $u$ implies that $F$ is strictly increasing at $c$, and Lemma 21.1(i,ii) in \cite{V98} shows that $F(c-\epsilon/2)<u\leq F(c)$. Thus $F(c-\epsilon/2)<u<F(c+\epsilon/2)$. The lemma also shows that $\hat{c}\leq c-\epsilon$ if and only if $\mathrm{P}(X_n\leq c-\epsilon\mid\mathcal F_n)\geq u$, and that $\hat{c}>c+\epsilon$ if and only if $\mathrm{P}(X_n\leq c+\epsilon\mid\mathcal F_n)<u$. It therefore suffices for us to show that
			\begin{equation}\label{eq:annoyinglemma}
				\mathrm{P}\left(\mathrm{P}(X_n\leq c-\epsilon\mid\mathcal F_n)\geq u\right)\to0\quad\text{and}\quad\mathrm{P}\left(\mathrm{P}(X_n\leq c+\epsilon\mid\mathcal F_n)<u\right)\to0.
			\end{equation}
			Let $h_1:\mathbb R\to[0,1]$ be a Lipschitz continuous function satisfying $\mathbbm 1(x\leq c-\epsilon)\leq h_1(x)\leq\mathbbm 1(x\leq c-\epsilon/2)$. Then
			\begin{equation*}
				\mathrm{P}(X_n\leq c-\epsilon\mid\mathcal F_n)\leq\mathrm{E}(h_1(X_n)\mid\mathcal F_n)\convp\mathrm{E}(h_1(X))\leq F(c-\epsilon/2)<u,
			\end{equation*}
			which establishes the first part of \eqref{eq:annoyinglemma}. A nearly identical argument using a Lipschitz continuous function $h_2:\mathbb R\to[0,1]$ satisfying $\mathbbm 1(x\leq c+\epsilon/2)\leq h_2(x)\leq\mathbbm 1(x\leq c+\epsilon)$ establishes the second part of \eqref{eq:annoyinglemma}.\quad\raisebox{-1pt}{$\Box$}
		\end{proof}
		
		\begin{proof}[Proof of \cref{prop:tauinfinity}]
			Let $Q_I$ be the quantile function for $\mathcal H'_{I}(\mathcal R)$, and let $c_{1-\alpha}=Q_I(1-\alpha)$. It follows from \cref{prop:bswc} and the definition of convergence in distribution conditional on the data in probability that
			\begin{equation*}
				\mathrm{E}\big(f(T_{n}^{1/2}(  \hat{R}^\ast-\hat R))\mid\{X_i^1\}_{i=1}^{n_1},\{X_i^2\}_{i=1}^{n_2}\big)\convp\mathrm{E}f(\mathcal R)
			\end{equation*}
			for every Lipschitz continuous map $f:L^1[0,1]\to[0,1]$. In particular, since $\mathcal H'_{I}:L^1[0,1]\to\mathbb R$ is Lipschitz continuous, we may choose $f$ to be the composition of an arbitrary Lipschitz continuous map $h:\mathbb R\to[0,1]$ with $\mathcal H'_{I}$. Recalling \eqref{eq:bSint2}, this yields
			\begin{equation*}
				\mathrm{E}\big(h(\hat{S}^\ast)\mid\{X_i^1\}_{i=1}^{n_1},\{X_i^2\}_{i=1}^{n_2}\big)\convp\mathrm{E}h\big(\mathcal H'_{I}(\mathcal R)\big).
			\end{equation*}
			It therefore follows from \cref{lem:bsQconsistent} that $\hat{c}_{1-\alpha}\convp c_{1-\alpha}$ if $Q_I$ is continuous at $1-\alpha$. Continuity of $Q_I$ is established by \cref{lem:nondegenerate}. Moreover, $c_{1-\alpha}>0$ because $\mathcal H'_I\geq0$ and because $Q_I$ is strictly increasing on $(1/2,1)$ by \cref{lem:nondegenerate}.
			
			Let $F_R$ and $Q_R$ be the cdf and quantile function for $\mathcal H'_R(\mathcal R)$, and let $b_{1-\alpha}=Q_R(1-\alpha)$. We have $b_{1-\alpha}\leq c_{1-\alpha}$ because $\mathcal H'_R\leq\mathcal H'_I$. Suppose that $\mathrm{H}_0$ is true. Then $\hat{S}\convd\mathcal H'_R(\mathcal R)$ by \cref{prop:F}. If $B_0$ has zero measure then $\mathcal H'_R(\mathcal R)=0$ and thus $\hat{S}\convp0$. Since $\hat{c}_{1-\alpha}\convp c_{1-\alpha}>0$, it follows that $\mathrm{P}(\hat{S}>\hat{c}_{1-\alpha})\to0$. On the other hand, if $B_0$ has positive measure then $F_R$ is continuous at $b_{1-\alpha}$ by \cref{lem:nondegenerate} and thus $F_R(b_{1-\alpha})=1-\alpha$ and
			\begin{equation*}
				\mathrm{P}(\hat S>\hat c_{1-\alpha})\leq\mathrm{P}(\hat S-(\hat{c}_{1-\alpha}-c_{1-\alpha})> b_{1-\alpha})\to1-F_R(b_{1-\alpha})=\alpha.
			\end{equation*}
			This proves assertion (ii) in \cref{prop:tauinfinity}. The last relation $\leq$ holds with equality if $F_1=F_2$ because in this case $R=I$ and $b_{1-\alpha}=c_{1-\alpha}$. Thus assertion (i) is also proved.
			
			If instead $\mathrm{H}_1$ is true, then $\mathrm{P}(\hat{S}>c_{1-\alpha}+1)\to1$ by \cref{prop:F}, and so
			\begin{equation*}
				\mathrm{P}(\hat{S}>\hat{c}_{1-\alpha})\geq\mathrm{P}(\hat{S}>c_{1-\alpha}+1)-\mathrm{P}(\hat{c}_{1-\alpha}\geq c_{1-\alpha}+1)\to1,
			\end{equation*}
			again using the fact that $\hat{c}_{1-\alpha}\convp c_{1-\alpha}$. This proves assertion (iii).\quad\raisebox{-1pt}{$\Box$}
		\end{proof}

		\begin{proof}[Proof of \cref{prop:modtest}]
			Assertion (iii) is correct because $\mathrm{P}(\hat{S}>\eta)\to1$ under $\mathrm{H}_1$ by \cref{prop:F}, $\mathrm{P}(\hat{S}>\hat{c}_{1-\alpha})\to1$ under $\mathrm{H}_1$ by \cref{prop:tauinfinity}(iii), and $\tilde{c}_{1-\alpha}\leq\hat{c}_{1-\alpha}$. Assertion (ii) is correct because if $\mathrm{H}_0$ is true and $B_0$ has zero measure then $\mathcal H'_R=0$ and thus $\hat{S}\convp0$ by \cref{prop:F}. It remains to prove assertion (i).
			
			Assume that $\mathrm{H}_0$ is true and $B_0$ has positive measure.	Let $F_R$, $Q_R$ and $b_{1-\alpha}$ be defined as in the proof of \cref{prop:tauinfinity}. \cref{lem:bsQconsistent,lem:FS} together establish that $\tilde{c}_{1-\alpha}\convp b_{1-\alpha}$ if $Q_R$ is continuous at $1-\alpha$. Continuity of $Q_R$ is established by \cref{lem:nondegenerate}. Since $\mathrm{H}_0$ is true and $B_0$ has positive measure, \cref{lem:nondegenerate} also establishes that $Q_R$ is strictly increasing on $(1/2,1)$. Thus $F_R$ is continuous at $b_{1-\alpha}$ and $F_R(b_{1-\alpha})=1-\alpha$. \cref{prop:F} establishes that $\hat{S}\convd\mathcal H'_R(\mathcal R)$. We therefore have $\mathrm{P}(\hat{S}>\tilde{c}_{1-\alpha})=\mathrm{P}(\hat{S}-(\tilde{c}_{1-\alpha}-b_{1-\alpha})>b_{1-\alpha})\to1-F_R(b_{1-\alpha})=\alpha$.\quad\raisebox{-1pt}{$\Box$}
		\end{proof}

	\end{document}